\def\year{2017}\relax
\documentclass[letterpaper]{article}
\usepackage{aaai17}
\usepackage{times}
\usepackage{helvet}
\usepackage{courier}
\usepackage{amssymb}
\usepackage{graphicx}
\usepackage{url}
\usepackage{graphicx}
\usepackage{amssymb}
\usepackage{arydshln}
\usepackage{multirow}
\usepackage{color}
\usepackage{url}
\usepackage{algorithmic}
\usepackage{algorithm}
\usepackage{subfigure}
\usepackage{wrapfig}
\usepackage{textpos}
\usepackage{pslatex}
\usepackage{amsmath}
\usepackage{booktabs}
\usepackage{threeparttable}
\newcommand{\tristar}{$^{\ast\ast\ast}$}
\newcommand{\bistar}{$^{\ast\ast}$}

\begin{document}
\title{Shocking the Crowd: The Effect of Censorship Shocks on Chinese Wikipedia}

\author{Ark Fangzhou Zhang\\University of Michigan\\arkzhang@umich.edu\\ \And 
Danielle Livneh\\University of Michigan\\danilvnh@umich.edu \And   
Ceren Budak \\University of Michigan\\cbudak@umich.edu \AND  
Lionel P. Robert Jr.\\University of Michigan\\lprobert@umich.edu \And  
Daniel M. Romero\\University of Michigan\\drom@umich.edu}
\maketitle
\begin{abstract}
Collaborative crowdsourcing has become a popular approach to organizing work across the globe. Being global also means being vulnerable to shocks -- unforeseen events that disrupt crowds -- that originate from any country. In this study, we examine changes in collaborative behavior of editors of Chinese Wikipedia that arise due to the 2005 government censorship in mainland China. Using the exogenous variation in the fraction of editors blocked across different articles due to the censorship, we examine the impact of reduction in group size, which we denote as the shock level, on three collaborative behavior measures: volume of activity, centralization, and conflict. We find that activity and conflict drop on articles that face a shock, whereas centralization increases. The impact of a shock on activity increases with shock level, whereas the impact on centralization and conflict is higher for moderate shock levels than for very small or very high shock levels. These findings provide support for threat rigidity theory -- originally introduced in the organizational theory literature -- in the context of large-scale collaborative crowds.

\end{abstract}

\section{Introduction}

Crowdsourcing is now poised to fundamentally transform the way we coordinate work \cite{anya2015bridge}. Online collaborative crowdsourcing platforms such as Wikipedia present a unique opportunity to tackle complex problems \cite{baldwin2010modeling,yu2011generating}. Shocks are unforeseen events that can disrupt and even threaten crowds \cite{cohendet2016always,jackson1988discerning,ocasio2011attention,staw1981threat}. Examples of such shocks include massive influxes or outflows of members, platform or government policies, or exogenous events (like the death of a celebrity) that increase the importance and visibility of the crowd's work. 

A number of studies have explored shocks and threats in organizations \cite{cohendet2016always,dutton2006explaining,Romero2016} as well as in small groups through experimental approaches  \cite{argote1989centralize,gladstein1985group,harrington2002threat}. However, less is known about how shocks affect online crowds, which often face distinct challenges to effectively respond to shocks \cite{robert2016influence,kittur2010beyond}. For example, online crowds have a much more fluid membership than offline groups and organizations, which makes a potential response to an unexpected shock much more difficult to organize. Therefore, a shock is likely to have a different impact on online crowds than their organizational work group counterparts. It is not clear how, or whether, online crowds respond to a shock. Part of the reason for this gap in the literature is the lack of adequate instances where the same type of shock affects a large number of online crowds -- a phenomenon that would allow for a systematic analysis of how crowds typically respond to shocks.

In this paper, we take a step to fill this knowledge gap by examining the impact of the 2005 Chinese government censorship block of Chinese Wikipedia. This event presents an exogenous shock to all Wikipedia articles that have contributors from mainland China, because these articles lose some and in some cases most of their contributors as a result of the censorship block. Using the exogenous variation in the fraction of editors blocked across different articles, we investigate the impact of shocks of different magnitudes on articles with varying numbers of editors. 

How do we expect the Chinese Wikipedia community to respond to this censorship shock? The literature on threat rigidity suggests that groups respond to an external threat or shock by centralizing their decision-making and decreasing internal conflict \cite{staw1981threat}. Should we expect Chinese Wikipedia crowds to behave like traditional offline groups? Our study aims to answer this question. We examine the crowds' response to shocks with respect to three collaborative behavior measures: activity, centralization, and conflict. 

Our main contributions are the following: \emph{(i)} We find that the overall activity level drops after a shock, but the exact drop in activity depends on the crowd's size; \emph{(ii)} as predicted by thread rigidity theory, centralization increases and conflict decreases when crowds are faced with moderate shocks. But surprisingly, the effects are less profound for more severe shocks; \emph{(iii)} our findings contribute to the organization theory literature by providing a large-scale validation of threat rigidity in a new emerging context; \emph{(iv)} our findings contribute to the crowdsourcing literature by providing analysis that could have important implications to the design and management of crowdsourcing platforms. 

\section{Related Work} 
{\bf \noindent Threat Rigidity and Centralization:}
Threat rigidity is often used to explain how groups behave when faced with an external shock \cite{kamphuis2011effects,staw1981threat}. This theory suggests that groups will seek to overcome external threats by increasing both the centralization of decision-making and group cohesion \cite{staw1981threat}. Centralization helps the group better coordinate its response to the threat during a time when coordination is difficult \cite{cohendet2016always}. Centralization also makes the group more efficient by leveraging its existing work practices while resources are low \cite{argote1989centralize}. Increases in cohesion reduce conflict \cite{windeler2015profiles}, which further facilitates coordination \cite{hinds2003out,cummings2009crossing}. Both increases in centralization and decreases in conflict allow the group to focus more on responding to the threat. 

Threat rigidity has been found to be consistent with behaviors observed in organizations \cite{cohendet2016always,dutton2006explaining} but less so in experimental studies of groups \cite{gladstein1985group,harrington2002threat,kamphuis2011effects}.
Other experimental studies found no evidence of centralization under threat \cite{argote1989centralize,driskell1991group}. 

There are several gaps in the current literature, which our study aims to fill. One, in the previously mentioned studies, due to the experimental methods employed, all threats were artificial. Our crowds are faced with a real external threat that could undermine their long term viability. Two, previous studies employed ad-hoc groups of people who had never worked together. Threats may not have much of an impact when members have no real history or future with their group. We overcome this limitation by examining crowds with both a history and a potential long term future. Three, group sizes have had little or no variance. This is particularly problematic given that previous research on crowds has shown that size is often related to both centralization and conflict \cite{arazy2011information,kittur2008harnessing}. We study crowds of different sizes which allows us to examine the interaction of crowd size and the effect of the shock on collaborative dynamics. Four, past studies examined the short-term impact of shocks on groups. Thus, even when these findings show a link between shocks and centralization it is difficult to know if such effects are lasting. In our study, we examined the impacts of shocks over a much longer period -- 1 year. Finally, other studies do not vary the intensity of the shock. Groups were either exposed to a shock or not exposed. In natural settings, shocks are likely to vary in intensity. In this study, shocks greatly vary in intensity, which allows us to examine their impacts over a range of levels.

{\bf \noindent Conflict and Crowds:} 
Coordinating work in large online crowds can be particularly difficult for several reasons. Unlike organizational work groups, online crowds often lack hierarchical structures, formal boundaries, stable memberships, and formal training \cite{kittur2010beyond,robert2015crowd,keegan2012editors}. Additionally, these crowds are typically composed of members who work at a distance and rely on electronic communication \cite{robert2016influence}, which further increases the prevalence of conflict \cite{hinds2003out,windeler2015profiles,filippova2016effects}. For example, \citeauthor{filippova2016effects} find that as task interdependence and geographic dispersion increase, so does conflict in Github crowds. \citeauthor{kittur2010beyond} examine  Wikipedia crowds and find that as crowd size increases coordination becomes more difficult and conflict increases. 

Other studies have focused on identifying ways to reduce conflict in online crowds. For example, \citeauthor{kittur2010beyond} discover that the positive relationship between crowd size and conflict diminishes when either communication between editors increases or crowds centralize their work. \citeauthor{filippova2016effects} find that leadership style and member participation in the decision-making reduce conflict in Github crowds. \citeauthor{arazy2011information} and \citeauthor{arazy2013stay} both find that crowds with more administrators have less conflict. 

This strand of literature offers an important and rich understanding of conflict and centralization in crowds. However, it has focused exclusively on conflict constructed under stable and static conditions. In online crowdsourcing platforms, crowds operate in environments that are far more chaotic and susceptible to disruptions than traditional organizational work groups. In this work, we aim to fill this gap by studying how centralization and conflict levels in collaborative online crowds change as a result of disruptive shocks. 

\section{Background}
Chinese Wikipedia, the Chinese-language version of Wikipedia, was established on October 24, 2002. As of October 2016, Chinese Wikipedia has accumulated over 4.8 million articles, with 43 million revisions contributed by over 2 million registered users. Aiming to provide a free online encyclopedia for Chinese-speaking users, Chinese Wikipedia has benefited from contribution by editors from mainland China (20.9\%), Hong Kong (26.3\%), Taiwan (36.9\%), the Unites States (5.6\%), and Canada (1.9\%).\footnote{\texttt{http://en.wikipedia.org/wiki/Chinese\_Wikipedia.}}

Due to the censorship of online content by the Great Firewall System of the Chinese government, Chinese Wikipedia had been blocked massively in mainland China three times by 2008. These blocks denied users from mainland China the access to Chinese Wikipedia. The first block took place on July 2, 2004, and was lifted on July 21, 2004. On Sep. 23, 2004, the Chinese government issued the second block, which lasted for 5 days. The third block of Chinese Wikipedia started on Oct. 19, 2005. Unlike the first two blocks, both of which lasted for only a short period of time, this block spanned for almost 1 year and was not lifted until Oct. 10, 2006 \cite{zhang2011}.

In this study, we focus on the impact of the shocks due to the third block on the collaborative behavior of editors of Chinese Wikipedia. There are two reasons for focusing on the third block. First, it was deployed without any prior announcement or warning and the Chinese government offered no official explanation afterward. Hence, it serves as an exogenous shock largely unexpected by editors of Chinese Wikipedia. Moreover, unlike the prior two blocks, this block spanned a relatively long period of time, which allows us to overcome the difficulty resulting from the overall sparsity of contribution to Wikipedia.

\label{sec:background}

\section{Identifying Blocked Users}

In order to study the effect of the block on collaborative groups of editors who maintain a specific article, we need to identify the blocked users. To provide a reliable identification of the blocked users, we make use of three criteria to decide whether a specific editor of Chinese Wikipedia is from mainland China and is therefore blocked during the censorship period of the third block: editing behavior, linguistic patterns, and temporal patterns.

{\bf \noindent Editing Behavior:} We first restrict the set of Wikipedia editors to those who made edits before the block. We then inspect the edits from these editors to filter out those who made edits during any of the three blocks -- those editors are either from outside mainland China, and thus unblocked, or have found methods to circumvent the censorship.

\begin{figure}[!h]
\includegraphics[scale=0.15]{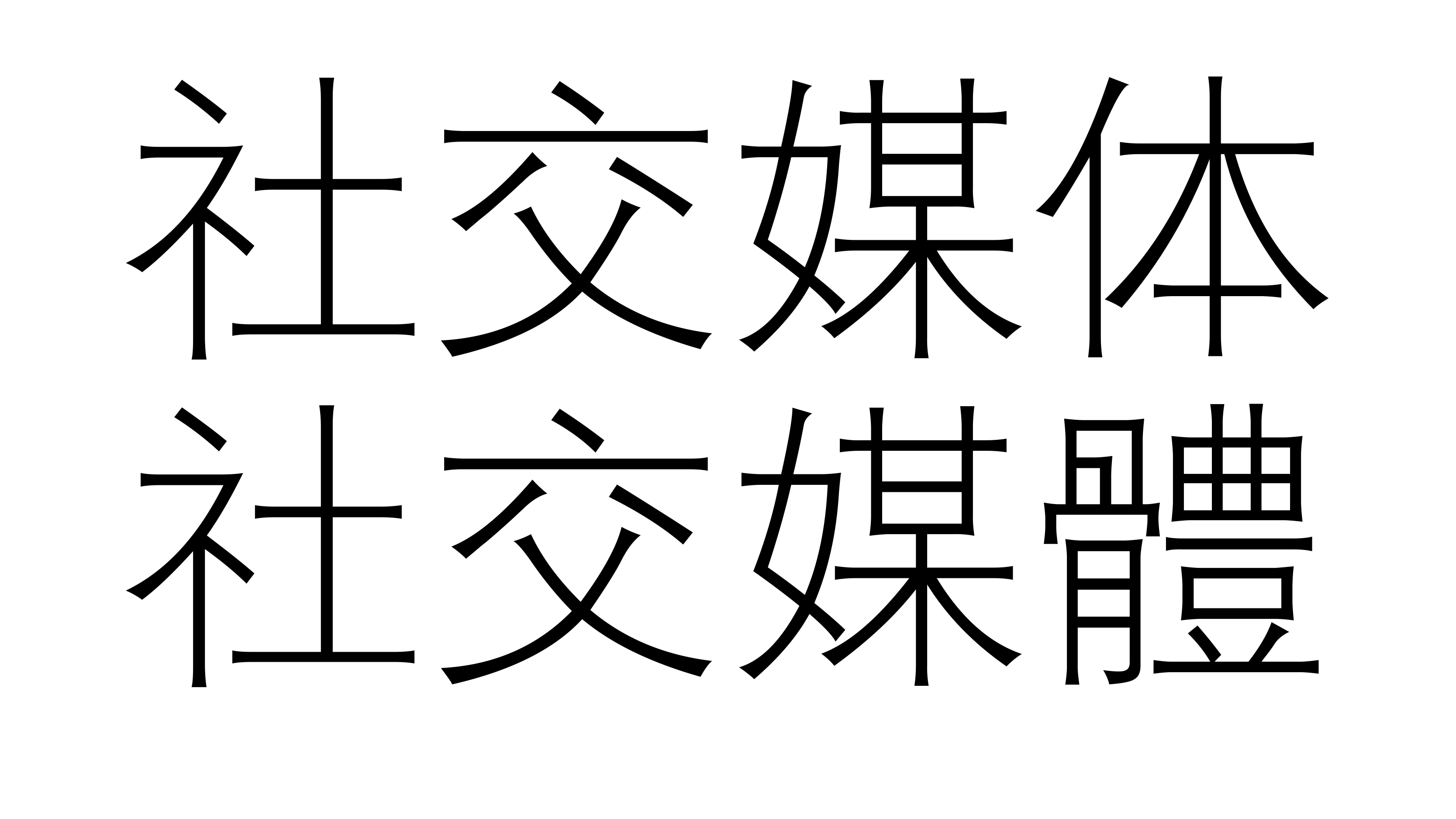}
\centering
\caption{Chinese characters for the word ``social media'': The first line is the simplified Chinese and the second line is the traditional Chinese.}
\label{fig:socialmedia}
\end{figure}

\begin{figure}
  \centering
  \includegraphics[scale=0.5]{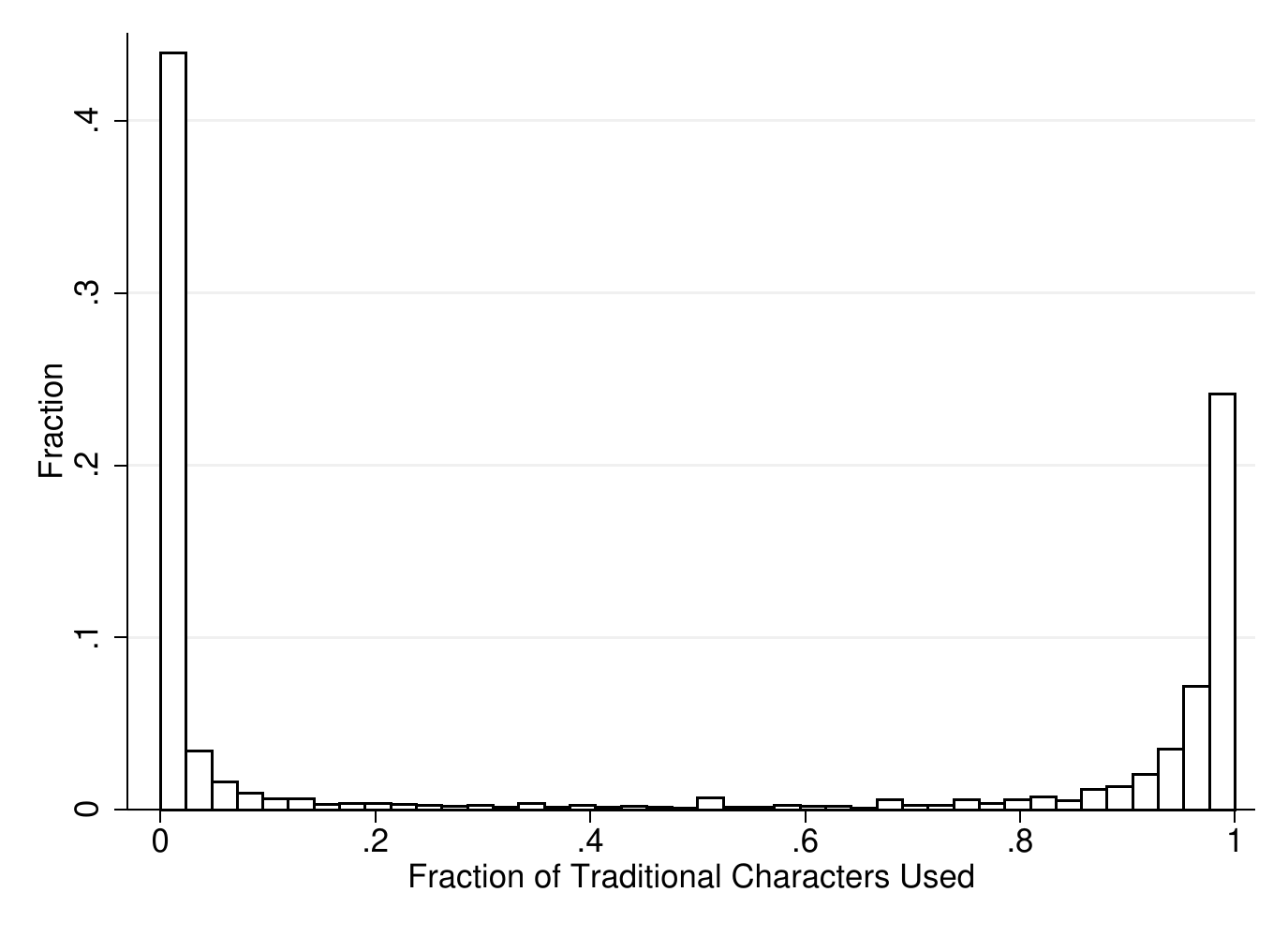}
  \caption[]{Traditional character usage: The y-axis denotes the fraction of the editors of Chinese Wikipedia that have x\% of their total characters written in traditional characters}
\label{fig:charUse}
\end{figure}

\begin{figure*} 
\centering
\subfigure[China]{
\includegraphics[scale=0.09]{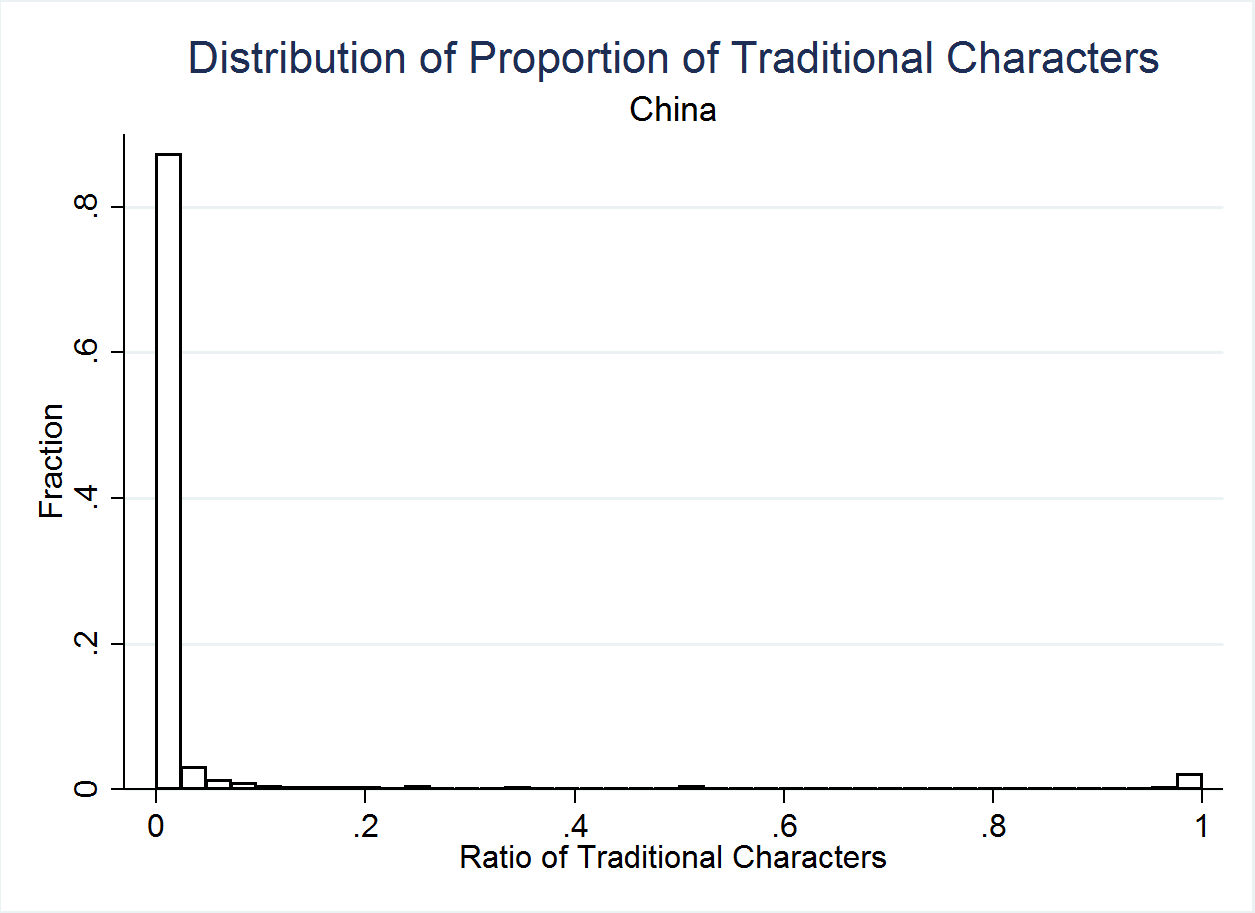}
\label{fig:chinaChar}
}
\subfigure[Hong Kong]{
\includegraphics[scale=0.09]{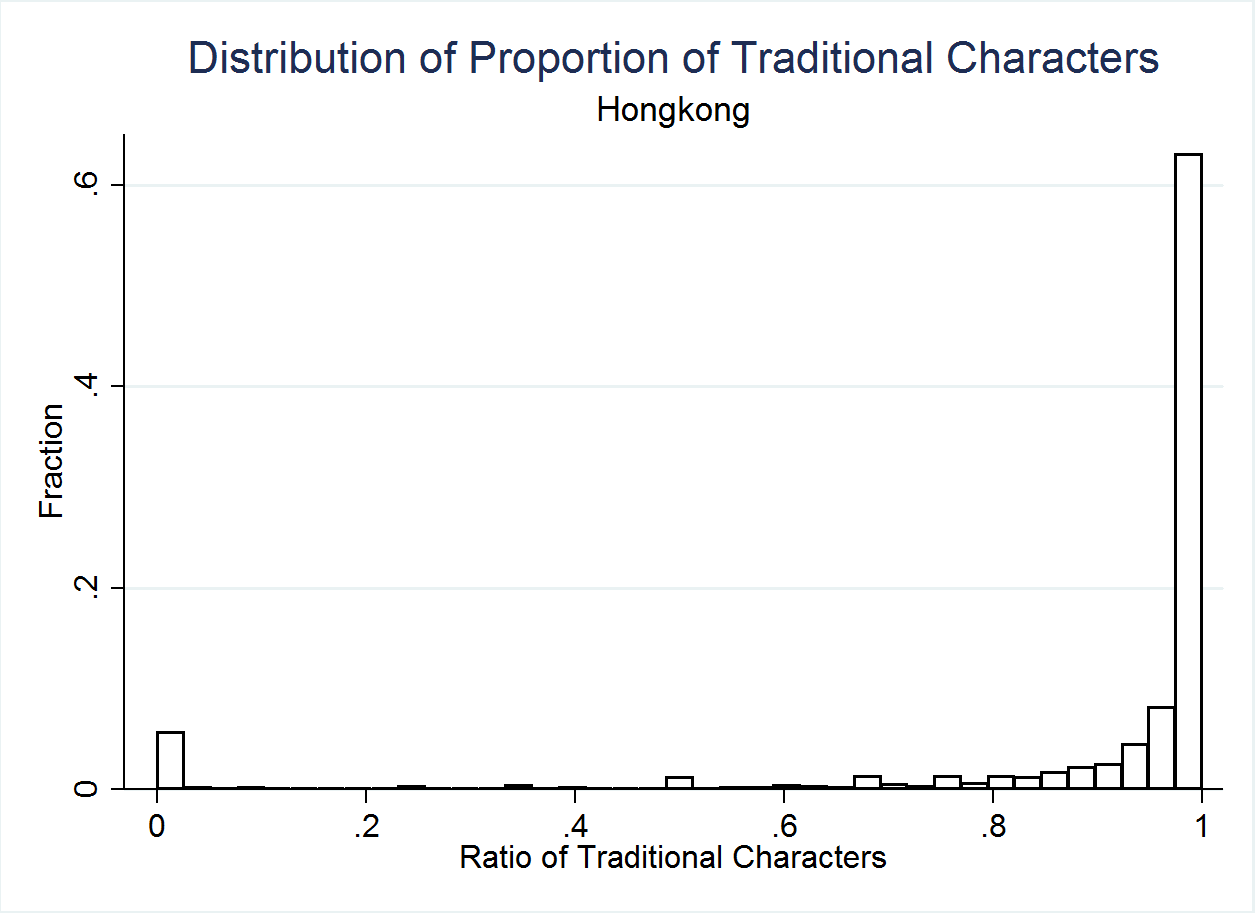}
\label{fig:hongkongChar}
}
\subfigure[Taiwan]{
\includegraphics[scale=0.09]{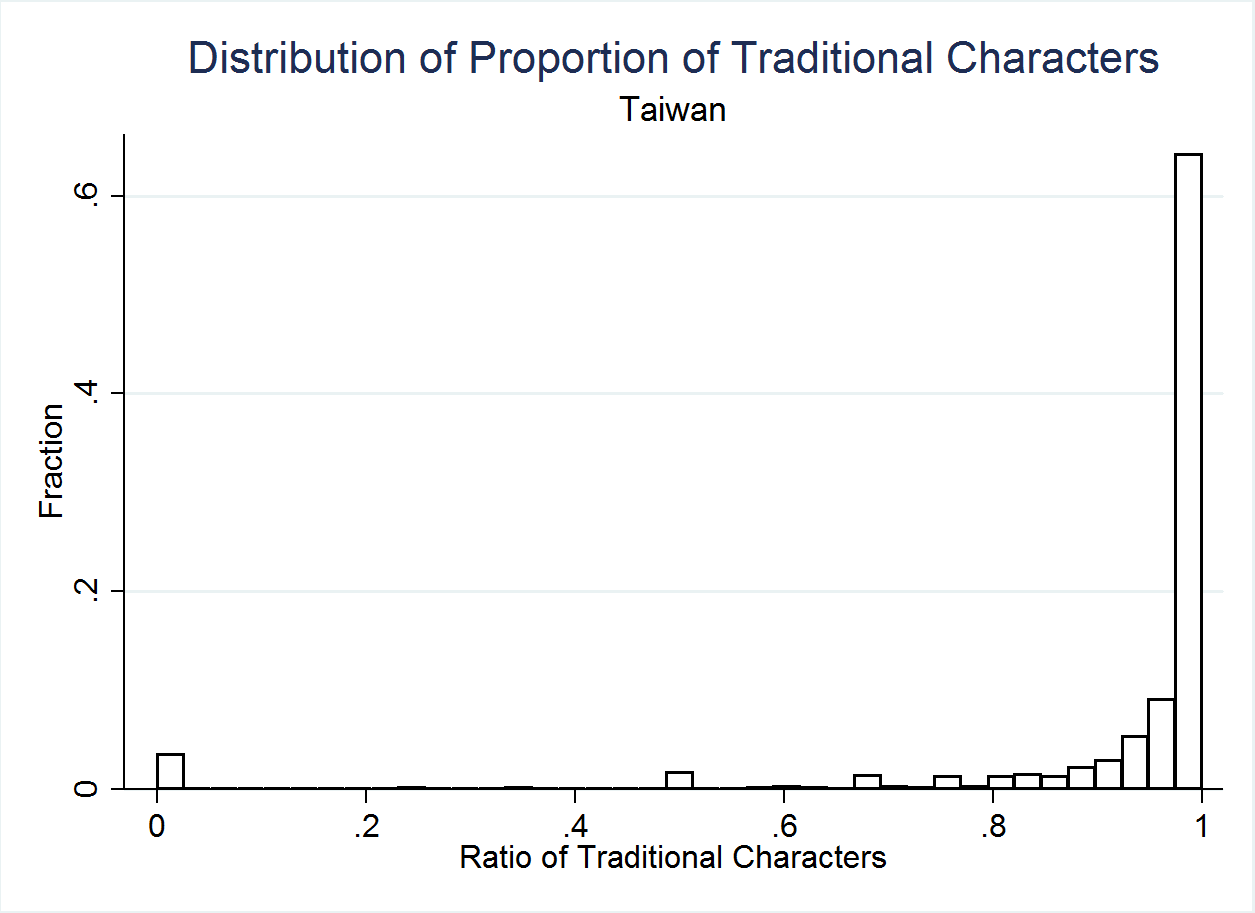}
\label{fig:taiwanChar}
}
\subfigure[U.S.]{
\includegraphics[scale=0.09]{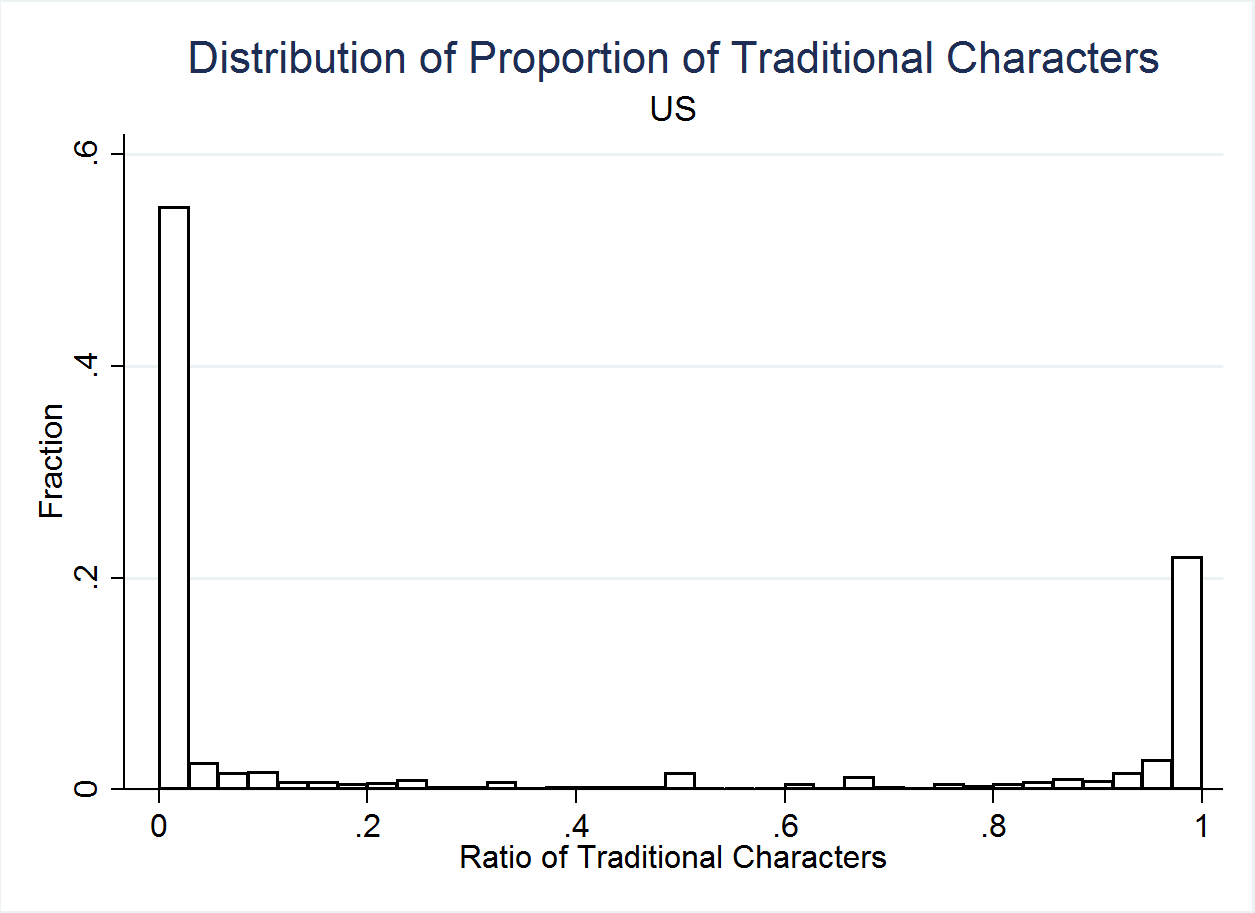}
\label{fig:usChar}
}
\caption{Traditional character usage across the four most common countries and areas}
\label{fig:charCountry}
\end{figure*}
{\bf \noindent Linguistic Patterns:} Our second check exploits the unique feature of the linguistic pattern of the Chinese language. There exist two encoding systems for Chinese language: the simplified Chinese and the traditional Chinese. Among all Chinese characters, there are approximately 2,000 for which the simplified version differs from the traditional version. For example, Figure \ref{fig:socialmedia} shows the Chinese characters for the word ``social media'' in the simplified version (the first line) and those in the traditional version (the second line). Note that the two versions share the first three characters, but differ from each other in the last one. The simplified Chinese is mainly used in mainland China, whereas the traditional Chinese is mainly used in Taiwan, Hong Kong and Macau. This feature provides a reliable measure for identifying mainland China editors, and has been used in related work \cite{zhang2011}. However, \cite{zhang2011} define an editor as non-blocked if more than 50 percent of the editor's additions are in traditional Chinese, a threshold that we ultimately find to be arbitrary. In comparison, we identify the optimal cut-off from the data. To motivate this point further, we present in Figure \ref{fig:charUse}: the distribution of editors in terms of their traditional character usage.\footnote{In this analysis we ignore characters that have the same representation under the two versions.} The plot reveals a prominent bimodal pattern -- editors consistently use either traditional or simplified Chinese encoding. In addition, we observe that the optimal cutoff lies closer to 20\%. Next, we demonstrate that the use of encodings does indeed vary across different countries in Figure \ref{fig:charCountry}. We produce this plot by considering edits from anonymous users, whose contributions are recorded by their IP addresses instead of usernames. This set of IP addresses allows us to map these editors and their use of encoding to their respective countries.\footnote{Anonymous users account for 78.89\% of all editors in our dataset, whose contribution represents no more than 13\% of all revisions prior to the third block.} The editors from mainland China consistently use the simplified version while those from Taiwan and Hong Kong use the traditional version. Given our results, we classify an editor as blocked only if 20\% or less of the characters used in their contributions are traditional characters.

\begin{figure}
  \centering
  \includegraphics[scale=0.33]{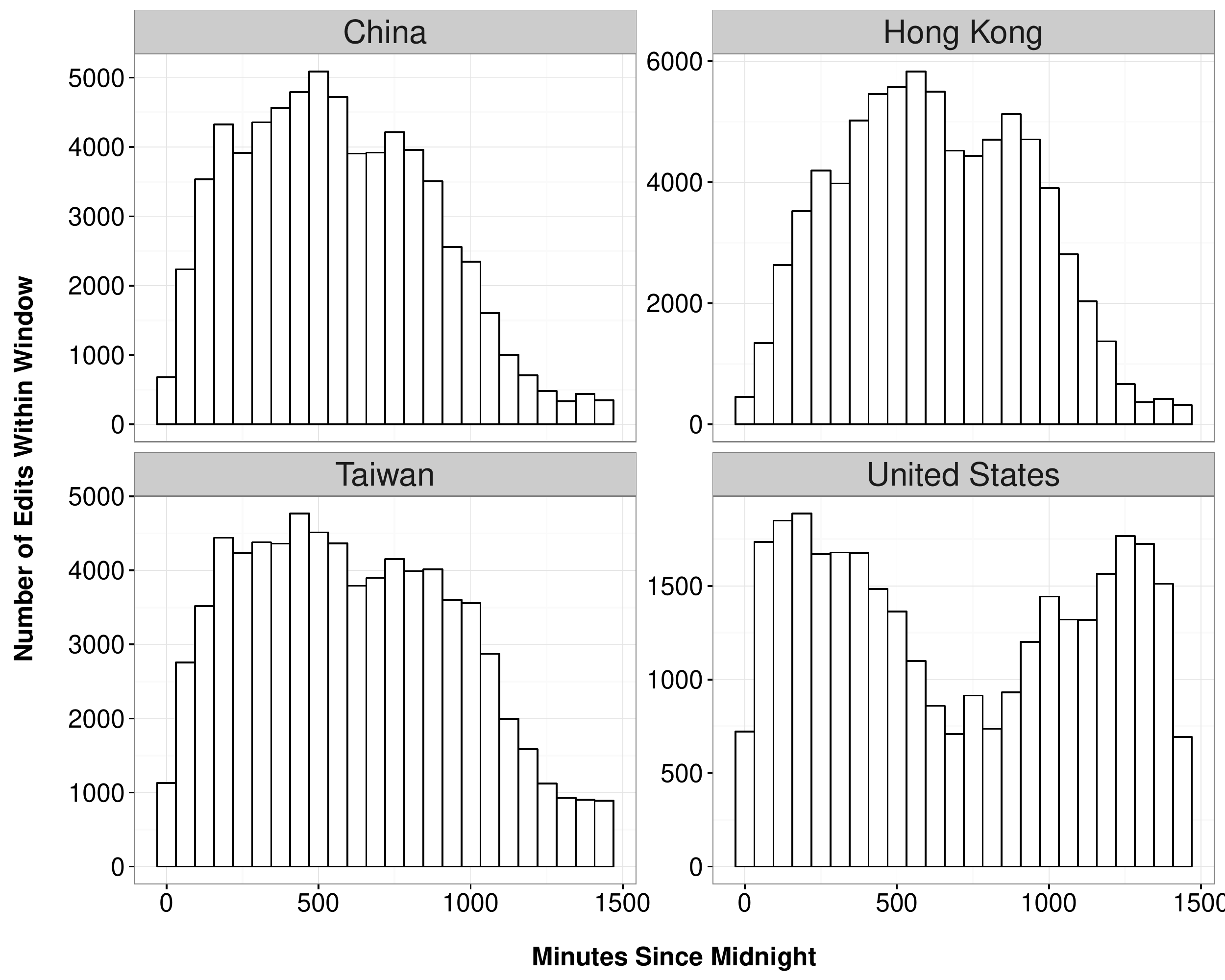}
  \caption{Daily temporal patterns in editing across countries and areas}
\label{fig:temporal}
\end{figure}

{\bf \noindent Temporal Patterns:} While Figure \ref{fig:charCountry} justifies the use of linguistic patterns to identify blocked users, Figure \ref{fig:usChar} also presents a challenge when considering the U.S. (or other countries with large Asian expat populations that are not included due to space limitations). We observe that while the U.S. population consists of both editors who use the simplified Chinese and editors that use the traditional Chinese, most of them use the simplified Chinese. Therefore the encoding technique might falsely classify a large number of editors from the U.S. as being from mainland China, where simplified encoding is also predominantly used. In an effort to remove these conflating editors, we also consider the daily editing patterns of editors from different countries (Figure \ref{fig:temporal}). We find that the editors in the U.S. contribute in different time frames from those in mainland China. We observe the sharpest difference for 18:00-24:00. Based on this finding, we classify an editor as being from mainland China, and therefore blocked, only if $y\%$ or less of their edits are contributed during this time frame.

In summary, we classify editors who pass the first test (described in editing behavior), use a traditional character at most $x\%$ of the time, and have at most $y\%$ of their edits contributed during the idle mainland China hours as blocked and the rest as unblocked.\footnote{We also attempted to differentiate between the truly blocked users and those who simply dropped out by fitting time lapses between two edits from a given editor to a Poisson distribution. Given the fitted distribution, we determine the likelihood of an editor to make a contribution during the block and identify an editor as blocked only if the likelihood of edit was above a threshold. This method does not improve the accuracy and therefore was not included in our final analysis.} Given that we have ground truth for 49,051 editors with IP addresses, we choose the values for these parameters that maximize the $F1$ measure when classifying this population. We find that the optimal $x=0.2$ and $y=1.0$. This setting results in $recall=1$, $precision=0.74$ and $F1=0.85$.

\section{Collaboration and Shock Measures}

Our goal is to characterize the effect that the block of Chinese Wikipedia in mainland China has on the dynamics of collaboration within an article. We consider each article as a unit of analysis and the set of Wikipedia users who edit the article as a collaborative crowd or team. We compare activity during the pre-block period (Oct. 19, 2004, to Oct. 19, 2005) and post-block period (Oct. 19, 2005, to Oct. 19, 2006). Because the number of editors and type of editors affected by the block varies across articles, we analyze the relationship between the fraction of edits made by blocked editors in an article and impact on three collaboration measures: level of activity, centralization of workload, and conflict. We now define these measures precisely. 

{\bf \noindent{Shock Level:}} Given an article $a$, we define the \emph{weighted blocked ratio} $B_{a}$ of article $a$ as the fraction of edits contributed by the editors blocked among all the edits during the pre-block period. This measure quantifies the intensity of the shock caused by the block on an article.

{\bf \noindent{Level of Activity:}} We first consider the effect that the block has on editing volume. For each article $a$, we let $EV_{a}^{\text{pre}}$ and $EV_{a}^{\text{post}}$ be the number of edits of $a$ made during the pre-block period and post-block period, respectively. We then measure the relative change in number of edits as $EV_{a}^{\Delta} = (EV_{a}^{\text{post}} - EV_{a}^{\text{pre}})/EV_{a}^{\text{pre}}$. Because collaborative crowds lose members due to the block, we intuitively expect a decrease in the total number of edits after the block. Previous literature suggests, however, that efforts to compensate for shock can perhaps prove effective \cite{kamphuis2011effects,sutcliffe2003organizing}. Thus, while it is unclear how the block may impact levels of activity, it is even less clear how activity levels interact with weighted blocked ratio. 

{\bf \noindent{Centralization:}} It is common for Wikipedia articles to have a skewed distribution of editors' contributions \cite{romero2015coordination}. Centralization is a form of coordination, where a few editors take charge of the majority of the work and rely on a large number of peripheral users to take on minor tasks, and make the crowd more effective by reducing the cost of explicit coordination \cite{kittur2008harnessing}. Meanwhile, a centralized article is less likely to be exposed to diverse expertise and points of view, which could limit the quality of the crowd's output \cite{arazy2006wisdom}. Overall, centralization can have an important impact on the coordination dynamics of the crowd and on the quality of its output.

To measure centralization, we use the Gini coefficient, a statistical measure of dispersion to quantify the level of inequality in a distribution \cite{dorfman1979formula}. We let $E^{\text{pre}}_{a}$ and $E^{\text{post}}_{a}$ be the set of editors of article $a$ in the pre-block and post-block period, respectively, and $N^{\text{pre}}_{a}$ and $N^{\text{post}}_{a}$ be the number of editors in $E^{\text{pre}}_{a}$ and $E^{\text{post}}_{a}$. We let $W^{\text{pre}}_a(e)$ and $W^{\text{post}}_a(e)$ be the number of times editor $e$ contributed to article $a$ in the respective time periods. We begin by computing the $G^{\text{pre}}_a$, the Gini coefficient of the set $\{\cup_{e \in E^{\text{pre}}_{a}} W^{\text{pre}}_a(e)\}$:
\begin{equation*}
G^{\text{pre}}_a = \frac{\sum_{i\in E^{\text{pre}}_{a}}\sum_{j\in E^{\text{pre}}_{a}} |e_i - e_j|}{2\sum_{i\in E^{\text{pre}}_{a}}\sum_{j\in E^{\text{pre}}_{a}} e_j}
\end{equation*}
For example, an article where every editor contributes the same number of edits has a Gini coefficient of 0, whereas an article with five editors who contribute 1 edit and one editor who contributes 20 edits has a Gini coefficient of 0.63. We calculate the corresponding Gini coefficient for the post-block period similarly. 

Because the value of Gini coefficient depends on the number of editors and edits in the article, we normalize $G^{\text{pre}}_a$ and $G^{\text{post}}_a$ by their maximum possible values given the number of editors and edits in article $a$ during the period for which we are calculating. We define the centralization of article $a$ during the pre-block period, $C^{\text{pre}}_a$,  as the fraction of $G^{\text{pre}}_a$ and the maximum value of $G_a$ given $E^{\text{pre}}_{a}$ and $EV^{\text{pre}}_{a}$. We also define the corresponding measures of centralization of an article during the post-block period in the same manner. Finally, we define the change in centralization as $C^{\Delta}_{a} = C^{\text{post}}_{a} - C^{\text{pre}}_{a}$. 

{\bf{Conflict:}} Wikipedia editors have access to a feature known as reverting that allows them to undo any other edit. When editors have disagreements with one another, they often engage in ``edit wars'', where they repeatedly revert one another's edits \cite{viegas2007talk,tsvetkova2016dynamics}. We use the fraction of edits that are reverts as a measure of conflict in an article during a given time period. We let $R^{\text{pre}}_{a}$ and $R^{\text{post}}_{a}$ be the number of reverts in article $a$ in the pre-block and post-block periods, respectively. We define the change in conflict in an article as $R^{\Delta}_{a} = R^{\text{post}}_{a} - R^{\text{pre}}_{a}$. 

\section{Results}

Here we present the results of our analysis on the change in activity, centralization, and conflict due to the block. For all subsequent analysis, we only consider articles with at least two editors before the block, as our goal is to understand how crowds respond to unexpected shocks. In addition, for all three measures we distinguish between the articles that have no editors from mainland China before the block from the articles that have at least one. These two populations are qualitatively different in important ways -- the former group does not appeal to a specific culture (mainland China). Indeed, articles from these two groups exhibit pre-block difference in characteristics that are relevant to our study. The articles that have at least one editor blocked have $9.06$ editors contributing $19.50$ revisions on average before the block, while those with no editors blocked have $4.45$ editors contributing $9.29$ revisions on average. In addition, the articles with at least one blocked editor tend to be contentious, on average exhibiting roughly a $50\%$ increase in rate of reverting compared to the group of articles with no blocked editors. 

Given these pre-block differences between the articles with and without any of their editors blocked, we provide two steps of analysis for each of our three measures. First, we compare the change in behavior between the articles with no editors blocked and the articles with at least one editor blocked. This allows us to understand whether being exposed to the shock, regardless of its level, has an effect on the articles. In total, we have 49,945 articles in our dataset and 27,856 among them have no editors blocked. Then, we examine how variations in the shock level affect articles that have at least one editor blocked. In short, we find that the shock negatively affects activity within groups, and that our findings for centralization and conflict are in agreement with literature on threat rigidity theory. 

\begin{table*}[t]
\centering
\begin{threeparttable}
\begin{tabular}{l c c c c c}
\toprule			
					& mean		& min 		& max		& std.dev	& skewness	 \\
\midrule
Activity			& -0.2927	& -1.0000	& 46.2857	& 1.0440	& 11.9280	\\
Centralization		& -0.0522	& -1.0000	& 1.0000	& 0.2169	& 0.2739	\\
Conflict			& -0.0235	& -0.6667	& 0.5000	& 0.1004	& 0.1827	\\
\bottomrule
\end{tabular}
\caption{\textit{Descriptive statistics of the collaboration measures.}}
\label{tab:desc}
\end{threeparttable}
\end{table*}

\subsection{Activity}
\label{sec:activityresults}

Table~\ref{tab:desc} provides descriptive statistics of the relative level of activity. An average article has a 29\% decrease in the level of activity, with the standard deviation of 1.044. We first compare the level of activity between articles with no editors blocked and those with at least one editor blocked. To this end, we regress the relative change in number of revisions of an article ($EV_{a}^{\Delta}$) over a dummy variable indicating whether or not the article has at least one editor blocked, denoted by $I_a^{Block}$, controlling for the number of editors, denoted by $N^{\text{pre}}_{a}$.
\begin{equation} \label{eq:activity_posblock}
EV_{a}^{\Delta} = \beta_0 + \beta_1 I_a^{Block} + \beta_2 N^{\text{pre}}_{a} + \epsilon_a
\end{equation}
Table \ref{tab:ttestresults} presents the result for regression \ref{eq:activity_posblock}. We see that articles on average become less active after the block, with a nearly 30\% decrease in the number of revisions. This might be explained by the reduction in the number of readers, which reduces individuals' incentives to contribute to the public good \cite{zhang2011}. Specifically, the volume of revisions for articles with no editor from mainland China shrinks by nearly 37\%. Compared with that, articles with at least one editor blocked experience an additional $3\%$ decrease (statistically significant at the 5\% level), which leads to a total 40\% decline in activity.\footnote{The variance inflation factor of the regression is 1.1, which indicates that the potential collinearity between the number of editors before the block and whether an article has any editor blocked does not severely affect the variance of the estimated coefficients.}

\begin{figure*} 
\centering
\subfigure[All Articles]{
\includegraphics[scale=0.4]{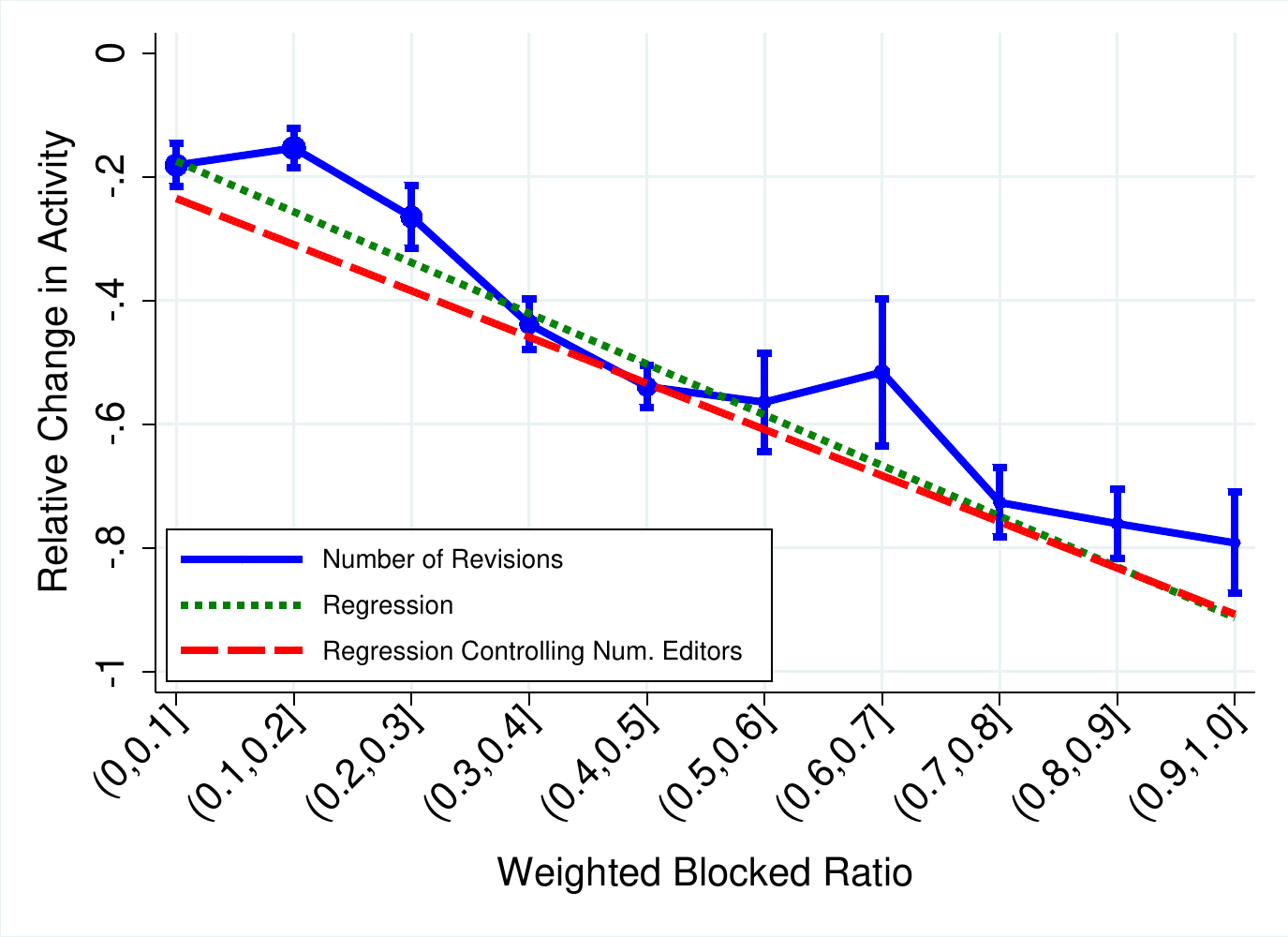}
\label{fig:resactivity}
}
\subfigure[Articles with $\leq 5$ editors]{
\includegraphics[scale=0.4]{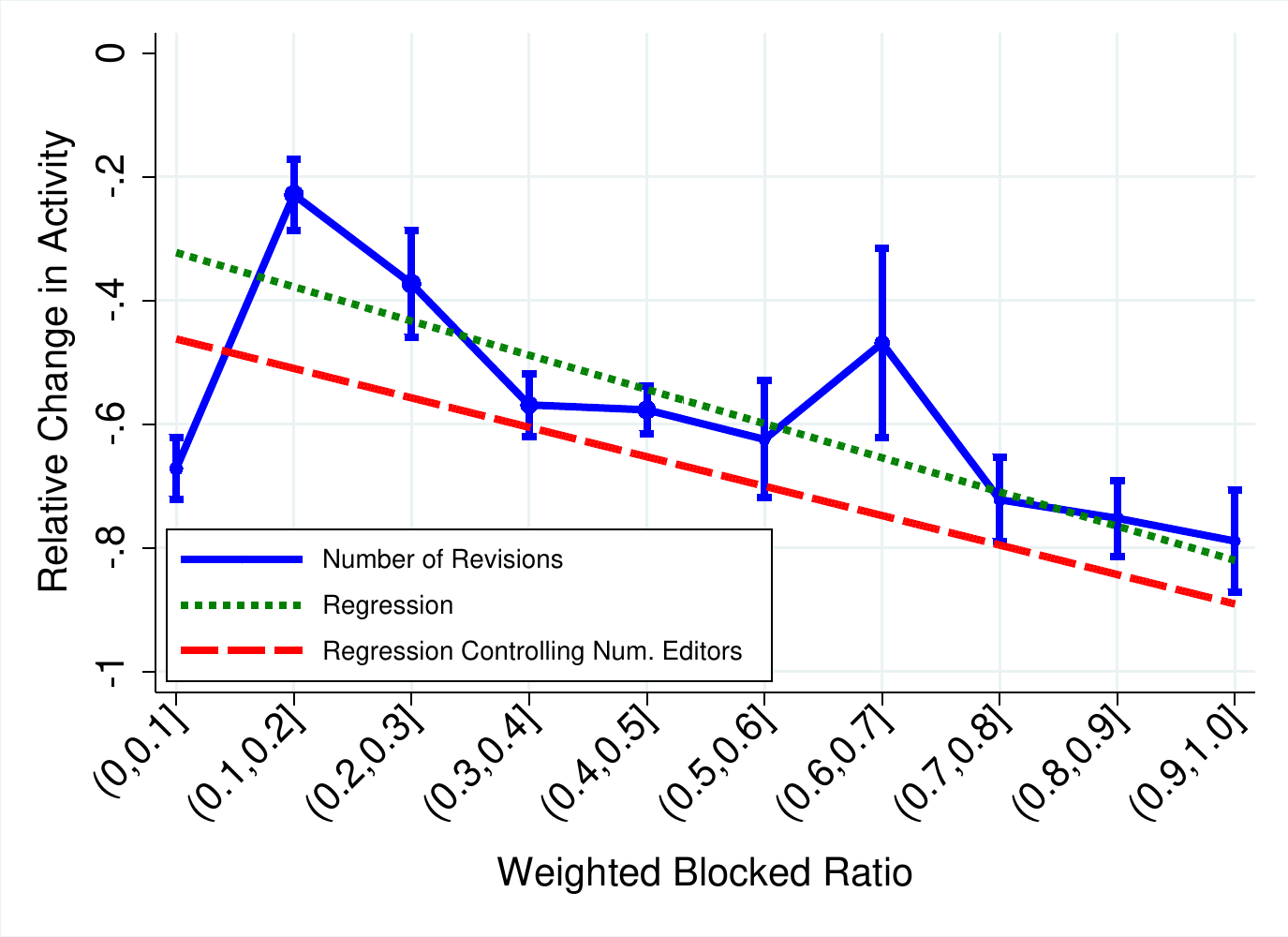}
\label{fig:resactivityLess}
}
\subfigure[Articles with $> 5$ editors]{
\includegraphics[scale=0.4]{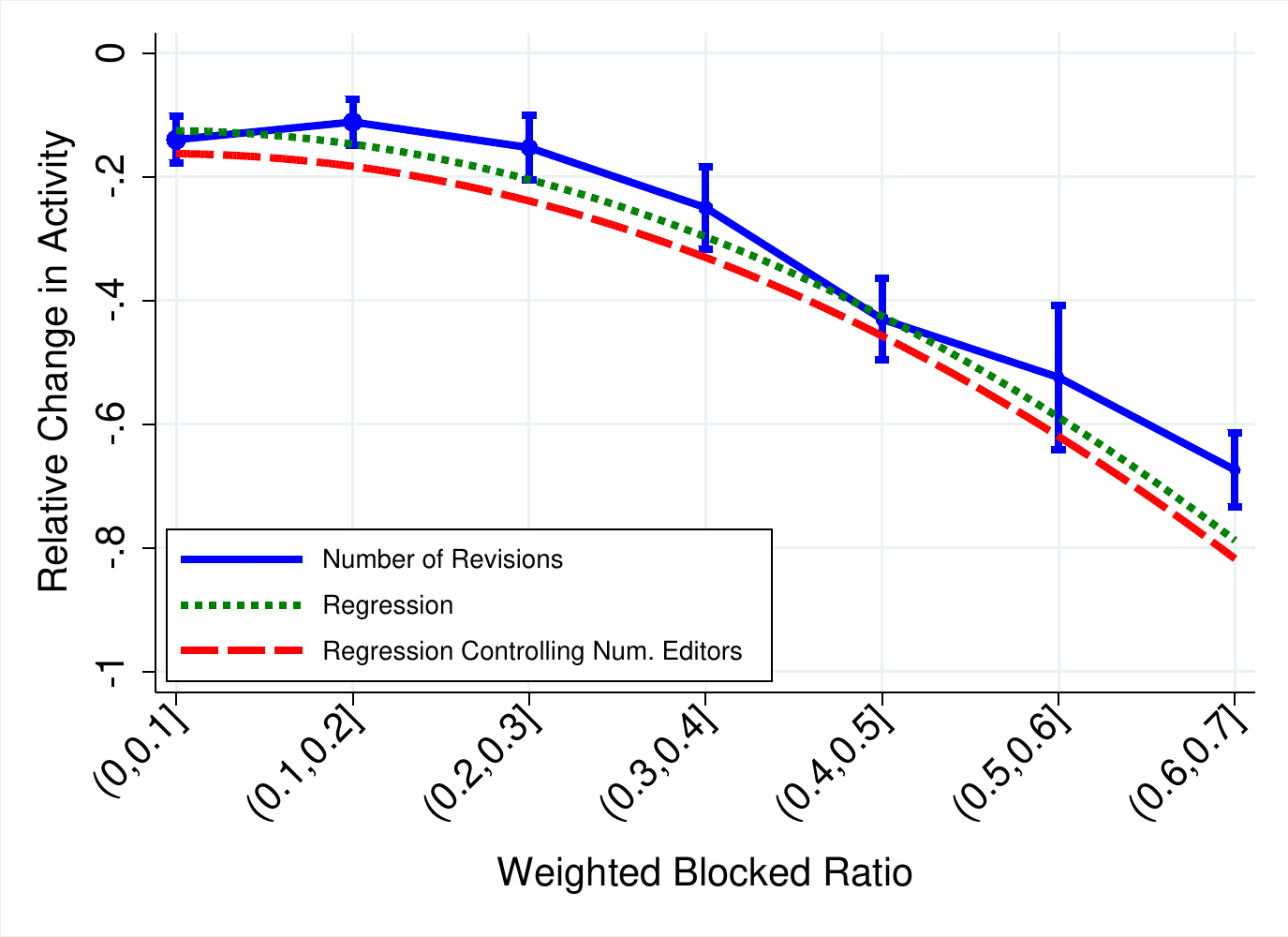}
\label{fig:resactivityGreater}
}
\caption{Change in editing volume as a function of shock level: The blue curve denotes the average $EV_{a}^{\Delta}$ within articles that were exposed to the shock level provided in the x axis. 1.96 standard errors are plotted for each point. The green hyphens indicate the regression fit without controlling for number of editors and the red dashes indicate the regression fit with control.}
\label{fig:resactivityAll}
\end{figure*}
Next, we examine the impact of various shock levels on the relative changes in activity across articles. To this end, we regress the relative change in number of revisions over the shock level (weighted blocked ratio) denoted by $B_a$, controlling for the number of editors before the block $N^{\text{pre}}_{a}$.
\begin{equation} \label{eq:activity_blocklevel}
EV_{a}^{\Delta} = f(B_{a}) + \beta N^{\text{pre}}_{a} + \epsilon_a
\end{equation}
In Table \ref{tab:results}, we report the regression results using both the linear specification and the quadratic specification for $f(B_{a})$.\footnote{Controlling for lifetime of articles in the regressions does not qualitatively affect the results.} To compare the fitness of the two models, for each specification we perform a likelihood ratio (LR) test of the quadratic model against the linear model. This shows that the quadratic model does not provide a significantly better fit to the change in activity than the linear model does.

Figure \ref{fig:resactivity} illustrates the relative change in activity due to various levels of shocks. The blue line plots the relative change in number of revisions as well as the error bars denoting 1.96 standard error above and below the mean. The red line and the green line represent the regression results with and without the number of editors before the block as a control. Hence, the discrepancy between the two regression lines illustrates how much of the change in articles' behavior is a result of the size of the crowd. We see from figure \ref{fig:resactivity} that articles subject to a higher level of shock experience a larger decrease in activity. Specifically, the regression results show that a 10\% loss in number of editors leads to a nearly 7.5\% decline in volume of revisions for an average article. We further separate the analysis between small crowds with at most 5 editors (Figure \ref{fig:resactivityLess}) and large crowds with over 5 editors (Figure \ref{fig:resactivityGreater}). Although the shock level still has a significant impact on both types of crowds, it appears to affect large crowds more substantially. Articles with more than 5 editors before the block exhibit non-linear decreases in number of revisions, whereas those with at most 5 editors respond to the shock in a linear way. Specifically, when the shock level is below 0.45, articles with more than 5 editors experience smaller changes in activity than those with no more than 5 editors do. When the shock level exceeds 0.45, the change in activity decreases faster in articles with more than 5 editors than those with less than 5 editors. This suggests that the vulnerability of a crowd to shocks depends on the size of the crowd. When the shock level is low, a large crowd is resilient toward it. However, beyond the threshold of 0.45, as the shock level increases, the change in the effect of a shock on activity level is more severe for large crowds.

\begin{table}[t]
\centering
\begin{threeparttable}
\begin{tabular}{l c c c}
\toprule
					& Activity 			& Centralization 	& Conflict 		\\
\midrule
$N^{\text{pre}}_{a}$			& 0.0119\bistar		& -0.0013\tristar	& -0.0001			\\
					& (0.0008)			& (0.0002)			& (0.0001)			\\
$I^{\text{Block}}$	& -0.0319\bistar	& 0.0007			& -0.0474\tristar	\\
					& (0.0127)			& (0.0029)			& (0.0030)			\\
constant			& -0.3691\tristar	& 0.0637\tristar	& 0.0253\tristar	\\
					& (0.0079)			& (0.0018)			& 0.0024			\\
\bottomrule
\end{tabular}
\begin{tablenotes}
\item The standard errors of the parameter estimates are provided in parentheses. \tristar and \bistar denote significance at 1\% and 5\%, respectively.
\end{tablenotes}
\caption{\textit{Regressions of (relative) change in activity, centralization and conflict over the dummy variable indicating whether an article has at least one editor blocked.}}
\label{tab:ttestresults}
\end{threeparttable}
\end{table}

\begin{table*}[t]
\centering
\begin{threeparttable}
\begin{tabular}{l c c c c c c c c c}
\toprule
			& \multicolumn{2}{c}{Activity}			&	& \multicolumn{2}{c}{Centralization}	&	& \multicolumn{2}{c}{Conflict} 	\\
            \cmidrule{1-3} \cmidrule{5-6} \cmidrule{8-9}
$E^{\text{pre}}$	& 0.0055\tristar	& 0.0055\tristar	&	& -0.0006\tristar	& -0.0005\tristar	& 	& 0.0002		& 0.0002			\\
			& (0.0008)			& (0.0008)			&	& (0.0002)			& (0.0002)			& 	& (0.0001)		& (0.0001)			\\
$B$			& -0.7471\tristar	& -0.7463\tristar	&  	& 0.1116\tristar	& 0.3916\tristar	& 	& 0.0013		& -0.2699\tristar	\\
			& (0.0478)			& (0.1472)			& 	& (0.0108)			& (0.0327)			& 	& (0.0105)		& (0.0293)			\\
$B^2$		& -					& -0.0011			& 	& -					& -0.3898\tristar	& 	& -				& 0.4430\tristar	\\
			&					& (0.1900)			&	&					& (0.0430)			&	&				& (0.0448)			\\
$\chi^2(1)$\tnote{$\dagger$}	&					& 0.00				&	& 					& 81.95				&	&				& 96.69				\\
LR test\tnote{$\dagger\dagger$}	&					& 0.9954			&	&					& 0.0000			&	&				& 0.0000			\\
\bottomrule
\end{tabular}
\begin{tablenotes}
\item[$\dagger$] Reports the test statistics for the likelihood ratio test.
\item[$\dagger\dagger$] Reports the $p$-value for the likelihood ratio test.
\item For each measure, the left column represents the result from the linear regression and the right column represents that from the quadratic regression. The standard errors of the parameter estimates are provided in parentheses. \tristar denotes significance at 1\%.
\end{tablenotes}
\caption{\textit{Regressions of the collaboration measures.}}
\label{tab:results}
\end{threeparttable}
\end{table*}

\subsection{Centralization}
\label{sec:centralizationresults}
We regress the change in centralization $C^{\Delta}_a$ over the indicator variable $I^{Block}_a$ controlling for the number of editors: 
\begin{equation}
C^{\Delta}_a = \beta_0 + \beta_1 I^{Block}_{a} + \beta_2 N^{\text{pre}}_a + \epsilon_a
\end{equation}

We find that the coefficient $\beta_1$ is not significantly different from zero, suggesting that there is no difference in the change in centralization between the two types of articles. To further investigate how the level of shock, $B_a$, relates to the changes in centralization  of articles with at least one editor blocked, we regress $C^{\Delta}_a$ over the shock level, $B_a$, controlling for the number of editors in the pre-block period:
\begin{equation}
C^{\Delta}_a = f(B_{a}) + \beta N^{\text{pre}}_a + \epsilon_a
\end{equation}

We fit both the linear and the quadratic models in the regression, and find that the quadratic one provides a significantly better fit than the linear model according to the likelihood ratio test ($p$-value $<$ 0.01\%).

\begin{figure*} 
\centering
\subfigure[All Articles]{
\includegraphics[scale=0.4]{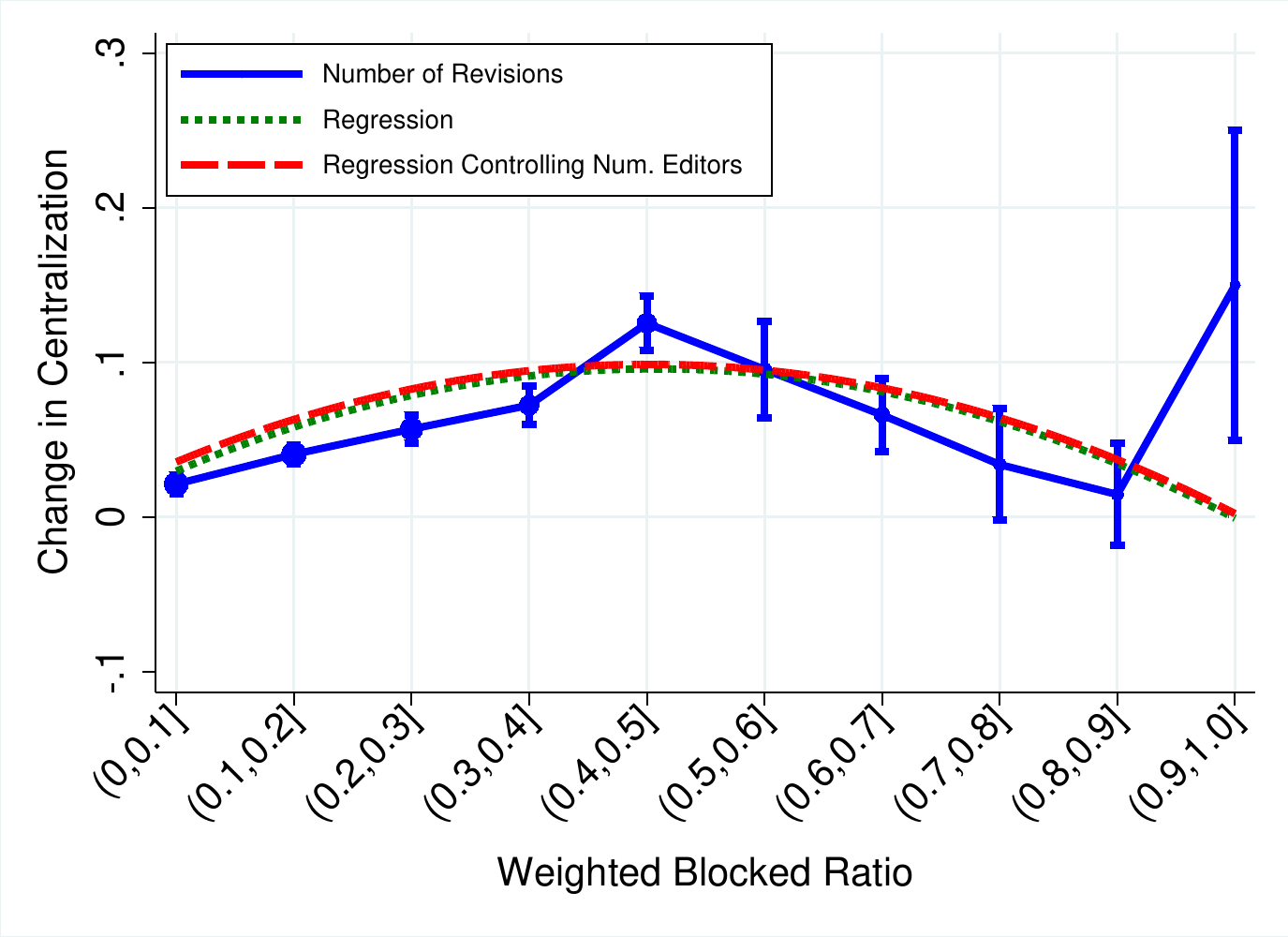}
\label{fig:centralizationAll}
}
\subfigure[Articles with $\leq 5$ editors]{
\includegraphics[scale=0.4]{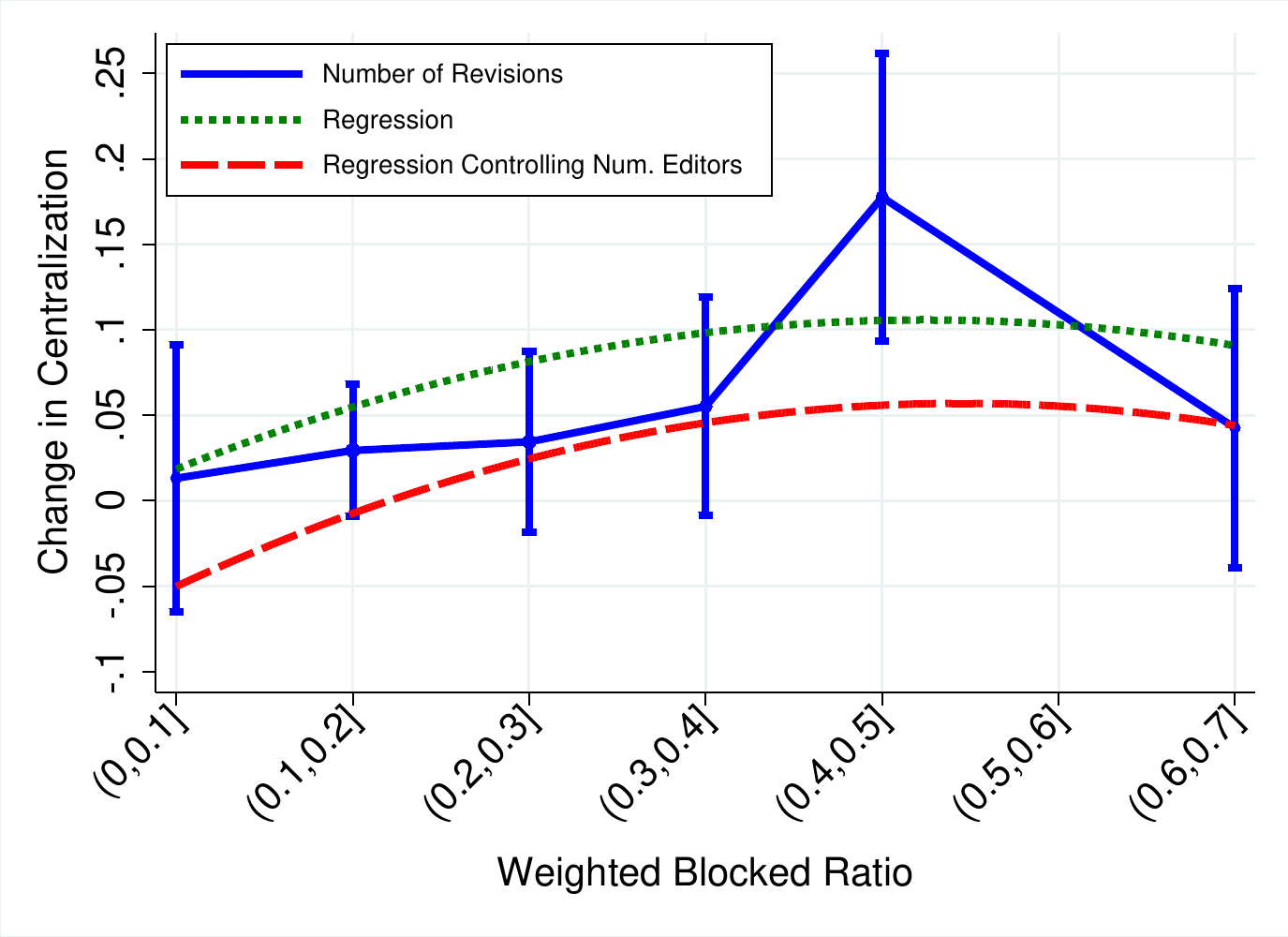}
\label{fig:centralizationLess}
}
\subfigure[Articles with $> 5$ editors]{
\includegraphics[scale=0.4]{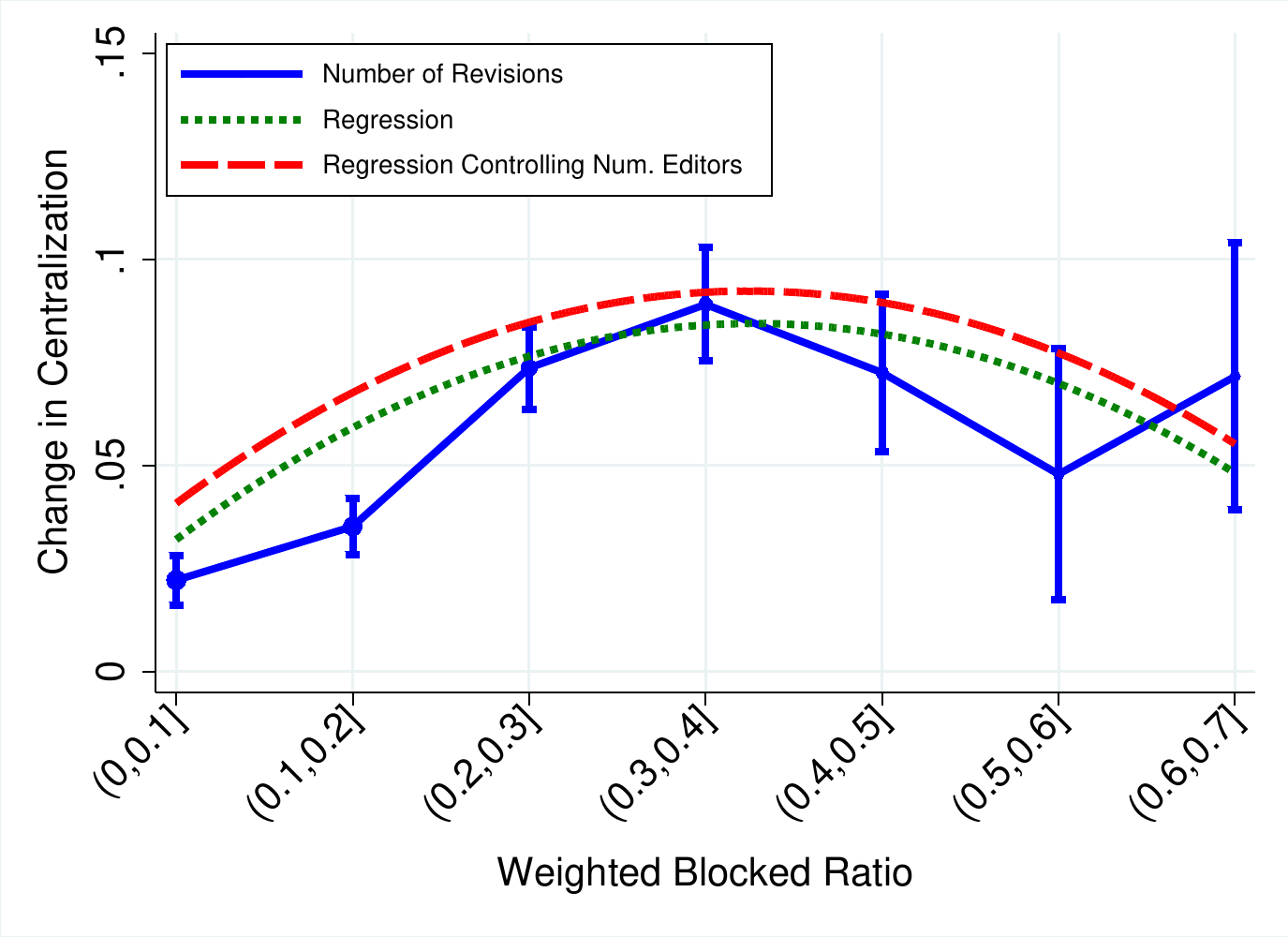}
\label{fig:centralizationGreater}
}
\caption{Change in centralization as a function of shock level.}
\label{fig:centralization}
\end{figure*}

\begin{figure}
  \centering
  \includegraphics[scale=0.4]{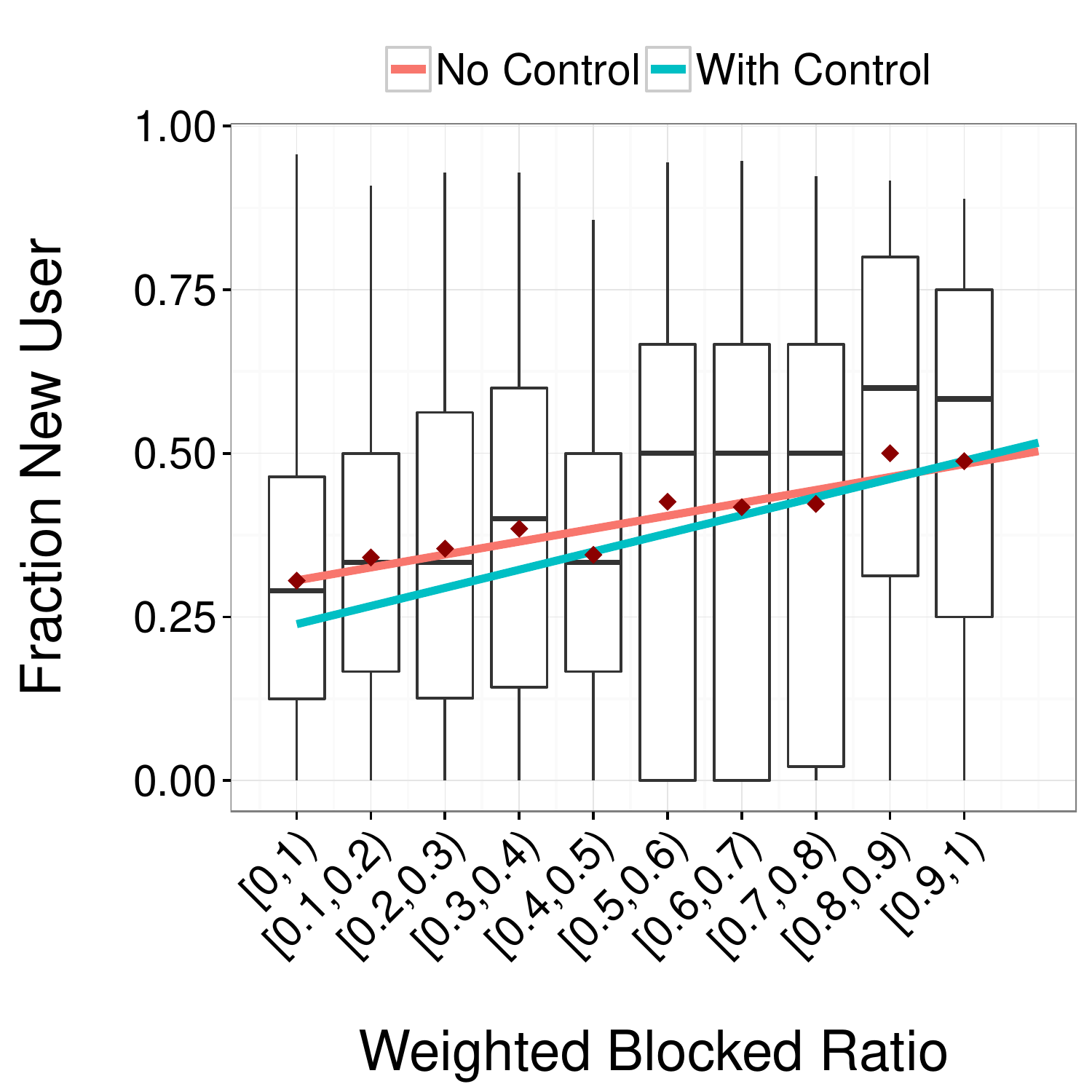}
  \caption{Change in composition as a function of shock levels.}
\label{fig:newUserBlock}
\end{figure}

Figure \ref{fig:centralization} illustrates the relationship between $C^{\Delta}_a$ and $B_{a}$ for all articles (a), articles with a small number of editors (b), and articles with a large number of editors (c). We find a consistent inverse U-shaped pattern -- that the change in centralization tends to increase with the shock level initially but decrease afterward. Using a back-of-the-envelope calculation based on the parameter estimates for the regression results in Table \ref{tab:results}, we find that the break-point at which the change in centralization begins to decrease is consistently around the point where $B_a = 0.5$ for the three sets of articles shown in Figure \ref{fig:centralization}.\footnote{The break point is given by $-\beta_B/2\beta_{B^2}$.}

The initial increase in centralization is consistent with the threat rigidity theory, which suggests that when groups have a perceived threat they become more centralized. So what explains the change in behavior beyond $B_a = 0.5$? We present the reasoning below.

We now investigate the change in composition of a group -- in terms of new versus old editors -- as a function of the shock level. We let $Comp_{a}^{\text{pre}}$ and $Comp_{a}^{\text{post}}$ be the number of editors who were active during the post-block period and who edited article $a$ for the first time during the pre-block period and post-block period, respectively. We then measure the fraction of new editors during the post-block period for an article $a$ as $Comp_{a} = \frac{Comp_{a}^{\text{post}}}{Comp_{a}^{\text{post}}+Comp_{a}^{\text{pre}}}$. In Figure \ref{fig:newUserBlock}, we show how this measure varies across articles with different shock levels. The x-axis in this figure denotes the shock level and the y-axis denotes the fraction of new users ($Comp_{a}$). We provide boxplots and means (diamond shape) for articles of varying shock levels. The red line and the green line are defined in the same manner as in the analysis of activity. The results show that as the shock level increases, the composition of a group post-block tends to include more new editors. This suggests that for the high shock levels, the composition of a group tends to be dominated by editors who joined the group later and therefore did not experience the shock. In fact, for $B_a >= 0.5$, the majority of editors for more than half of the articles are new to the group. 

Given the finding on compositional effects, the break point in Figure \ref{fig:centralization} is now easier to interpret. As we move beyond $B_a \ge 0.5$, for instance, the majority of a group consists of users who joined the article after the shock and thus did not experience the shock. It is natural that for such articles, the changes in concentration are not as strong as in the cases where most group members experienced the shock and hence behave according to threat rigidity theory.  

\subsection{Conflict}
\label{sec:conflictresults}
We analyze conflict by comparing articles with and without editors blocked in the following regression:
\begin{equation}
R^{\Delta}_a = \beta_0 + \beta_1 I^{Block}_{a} + \beta_2 N^{\text{pre}}_a + \epsilon_a
\end{equation}
We find that articles with at least one editor blocked experience a $2.2\%$ drop in conflict, while those with no editors blocked experience a $2.5\%$ increase in conflict. Given that articles with no editors blocked are not directly affected by the shock, this poses a conundrum. However, this can be explained once the trend of conflict is estimated during the pre-block period. Indeed, we find that articles with no editors blocked already experience a $2.5\%$ increase when comparing time periods October 2004-May 2005 to May 2005-October 2005, while those with at least one editor blocked experience $< 0.001\%$ change in the same time period. This shows that the trend in conflict is unchanged for articles with no editors blocked, while those with at least one editor blocked shifts from constant conflict to a decreasing one.

\begin{figure*} 
\centering
\subfigure[All Articles]{
\includegraphics[scale=0.4]{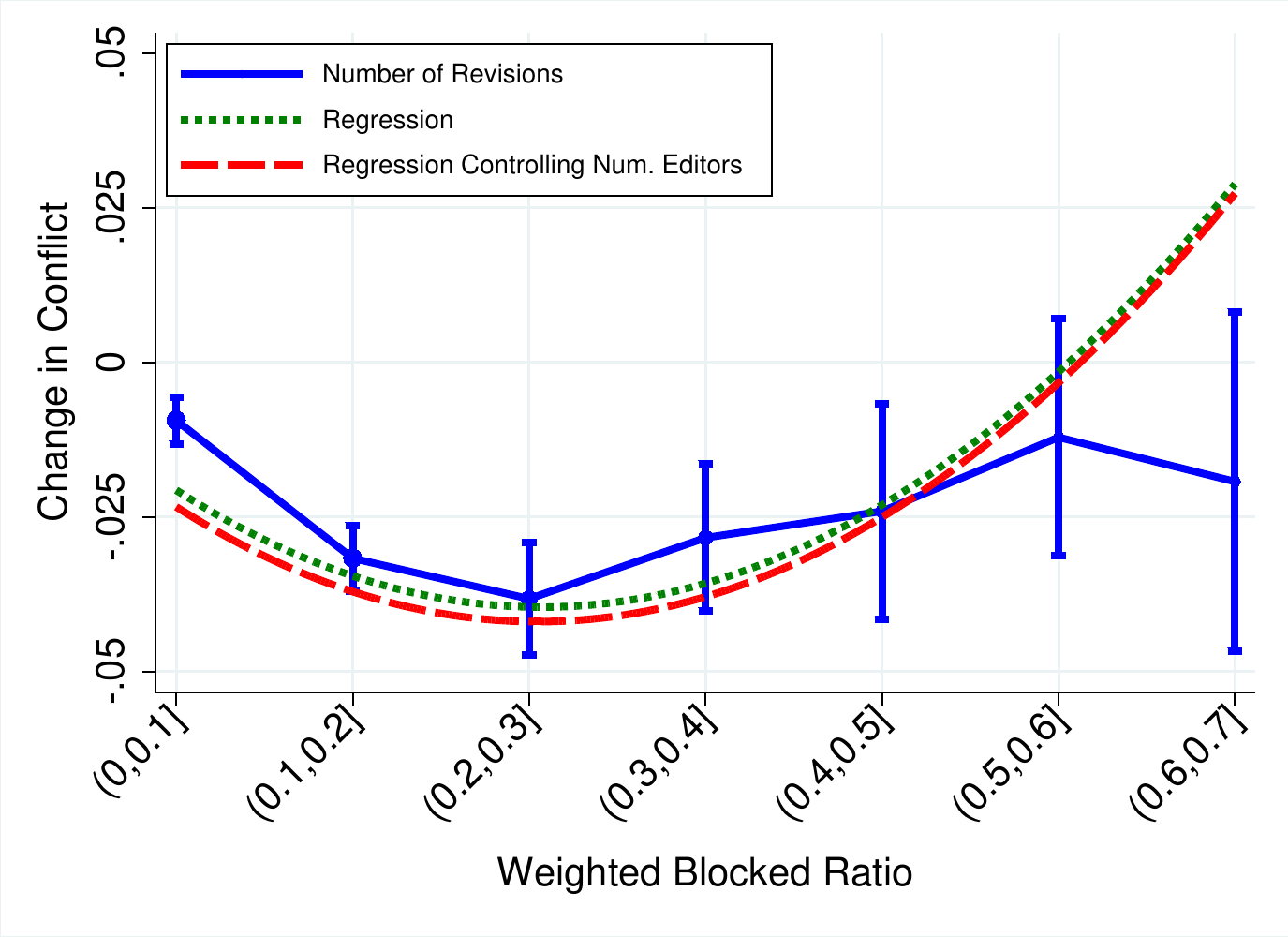}
\label{fig:conflictAll}
}
\subfigure[Articles with $\leq 5$ editors]{
\includegraphics[scale=0.4]{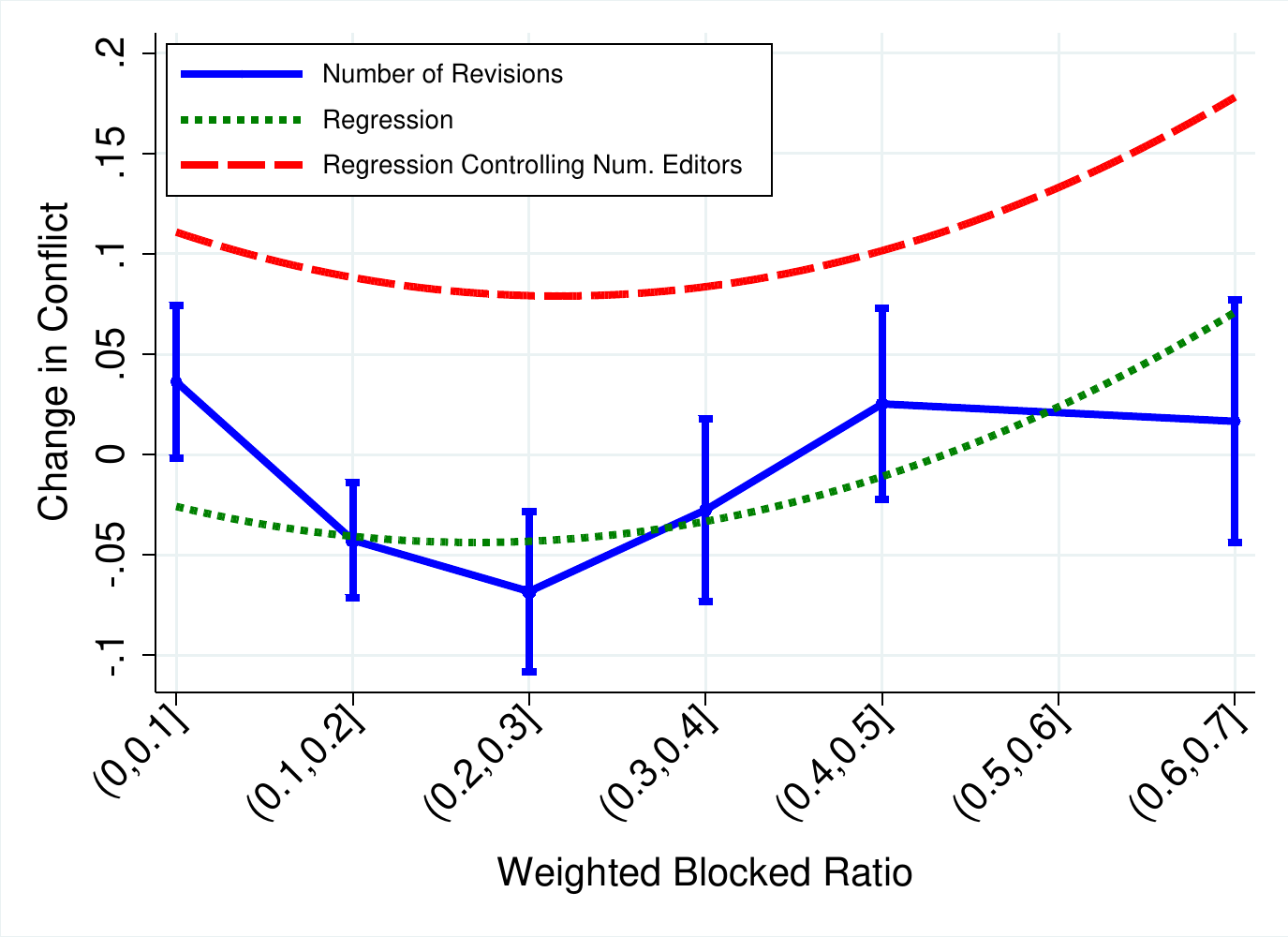}
\label{fig:conflictLess}
}
\subfigure[Articles with $> 5$ editors]{
\includegraphics[scale=0.4]{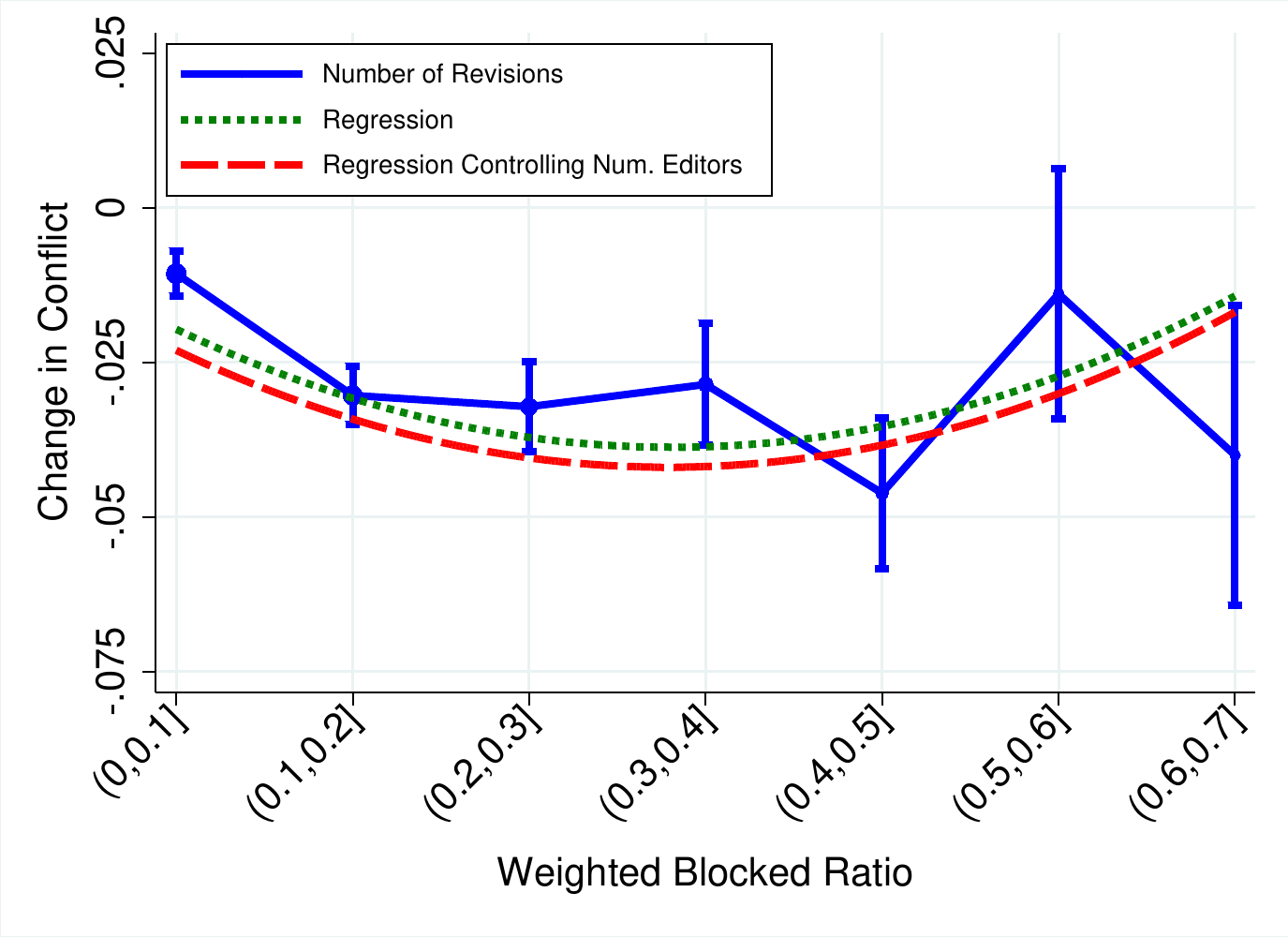}
\label{fig:conflictGreater}
}
\caption{Change in conflict as a function of shock level.}
\label{fig:conflict}
\end{figure*}

Next, we examine the effect of the shock level on conflict in articles. To that end, we regress the amount of conflict of an article to the weighted ratio of blocked editors, controlling for the number of editors as follows:
\begin{equation}
R^{\Delta}_{a} = f(B_{a}) + \beta N^{\text{pre}}_a + \epsilon_a
\end{equation}
Here, we consider articles that had at least one revert either one year before or after the block as this limits the analysis to articles with editors who are aware of the reverting feature. The results are consistent with the overall qualitative findings if all articles are included.
Note that we also evaluated the fit of a linear model and find that the quadratic model provides a significantly better fit for the data given the likelihood ratio test ($p$-value $<$ 0.01\%). The findings presented in Figure \ref{fig:conflict} are intriguing. We observe that for both small and large articles, small shocks result in a decrease in conflict. This finding is in agreement with threat rigidity theory, which suggests that when groups face an external threat they become more cohesive and hence exhibit less conflict \cite{staw1981threat}. 

Moving from small to big shocks, we find an inflection point -- after around $B_a > 0.3$, a larger blocked ratio results in a smaller reduction in conflict. As shown in the analysis of centralization, when large shocks occur, most of the current group members disappear and the new composition of the team consists of new group members. Thus, the increase in the change in conflict when $B_a$ is large is likely due to the fact that most group members in these cases did not experience the shock.

\subsection{Mediation Analysis}
Centralization and conflict are likely to relate to each other. When a group is highly centralized, explicit coordination is less costly and it is easier to complete tasks without engaging in conflict \cite{kittur2010beyond}. We observe that shock level had the opposite relationship with centralization than it does with conflict. Indeed, controlling for the shock level and the number of editors, we find a negative and significant relationship between centralization and conflict. It is possible that the shock affects conflict indirectly through its impact on centralization. To separate out the direct effect of weighted blocked ratio on conflict and any indirect effect through centralization, we conduct a mediation analysis \cite{mackinnon2007mediation} among weighted blocked ratio, centralization, and conflict.

\begin{figure}
  \centering
  \includegraphics[scale=0.3]{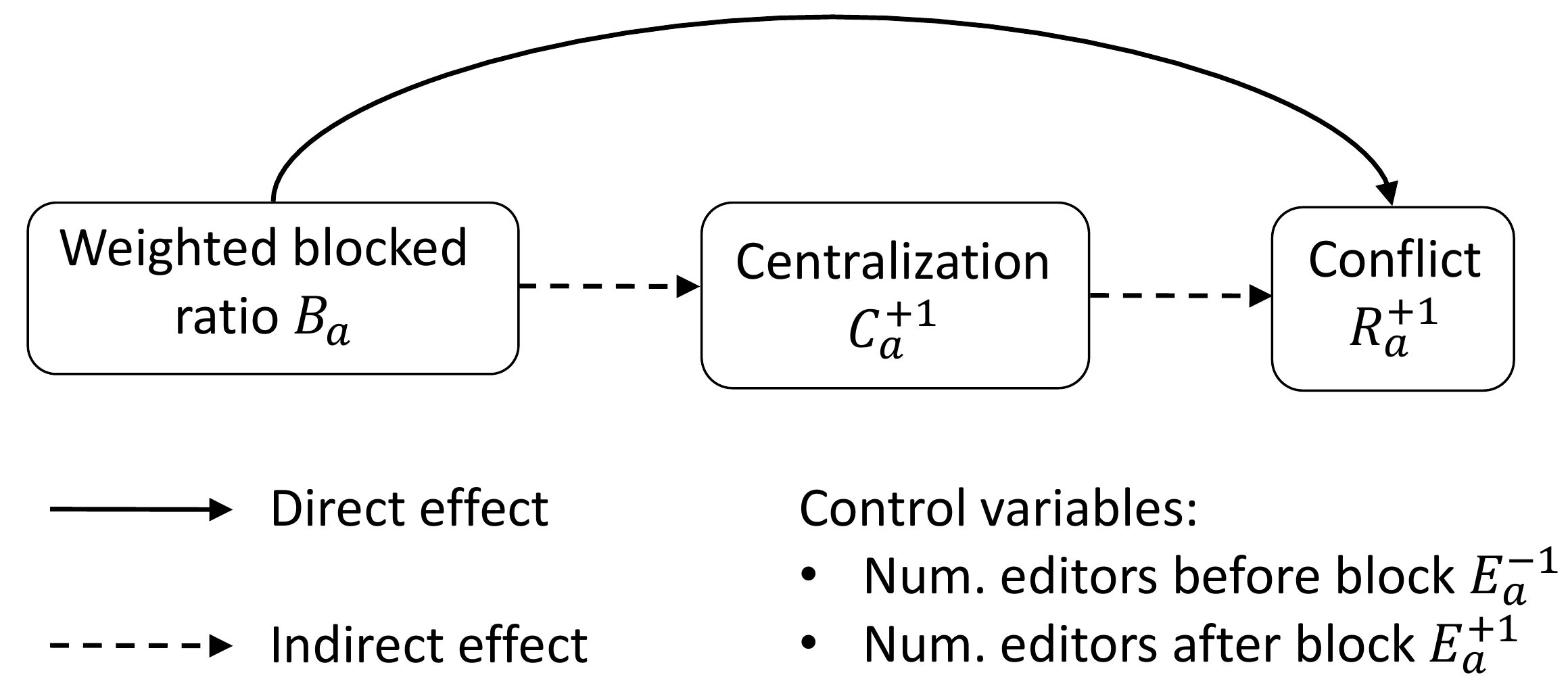}
  \caption[]{Mediation analysis diagram}
\label{fig:mediation_diagram}
\end{figure}

Figure \ref{fig:mediation_diagram} shows the model and Table \ref{tab:mediationanalysis} summarizes the decomposition of the direct effect and the indirect effect from the mediation analysis. The only significant effect that weighted blocked ratio has on conflict is the direct effect and there is no significant indirect effect through concentration. Indeed, the direct effect accounts for over 99\% of the total effect that the weighted blocked ratio has on conflict. This suggests that while centralization directly impacts conflict, the observed non-linear effect that weighted blocked ratio has on conflict is independent of the effect of centralization. 

\begin{table}[t]
\centering
\begin{tabular}{l c c}
\toprule
					& $B^2$ 			& $B$				\\
\midrule
Direct Effect		& 0.4222\tristar	& -0.2471\tristar	\\
					& (0.0448)			& (0.0295)			\\
Indirect Effect		& 0.0012			& -0.0011			\\
					& (0.0035)			& (0.0035)			\\
Total Effect		& 0.4234\tristar	& -0.2482\tristar	\\
					& (0.0446)			& (0.0293)			\\
\bottomrule
\end{tabular}
\caption{Results from the mediation analysis. The standard errors of the parameter estimates are provided in parentheses. \tristar denotes $p$-value $<0.01$ respectively.}
\label{tab:mediationanalysis}
\end{table}

\section{Discussion}
Through this research we seek to understand the impact of external shocks on crowds. To do so, we examine the 2005 Chinese government censorship of Wikipedia. Results from our analysis provide four overarching findings, which have implications for research and design.

First, group size matters. Although size is not a key element in the threat rigidity literature on groups, it had an important role in our study. Larger crowds were able to maintain similar levels of activity when they experienced moderate shocks. Smaller crowds experienced more dramatic drop-offs in their level of activity. This supports the idea of resiliency through size. However, the opposite was true when shocks were more severe. For severe shocks, smaller crowds experienced smaller decreases in their level of activity, while larger crowds had dramatic drop-offs in their activity. The importance of size in understanding how groups respond to shocks may have been de-emphasized in prior literature, which did not significantly vary group size. However, our results suggest that size is vital to understanding how groups respond to threats. 

Second, in the context of crowds the impact of shocks on centralization and conflict is not as straightforward as the literature suggests. Surprisingly, moderate shocks had a much more profound and lasting impact than severe shocks.  In cases of severe shocks large portions of the crowd were lost and later replaced with newcomers. The greater the influx of newcomers into the crowd, the less the crowd displayed evidence of the shock. More specifically, these crowds are more decentralized and have more conflict compared to crowds that experience more moderate shocks and retain more of their previous members. Newcomers did not experience the shock and are likely to be less willing to support increases in centralization and decreases in conflict. Although this finding is novel, it is unclear whether it only applies to crowds or it could generalize to other settings. 

Third, this study extends research on threat rigidity to include a large-scale validation in the context of online groups. As predicted by threat rigidity, crowds become more centralized and conflict decreased after those crowds experience a moderate shock. The fact that these are real groups and face a genuine threat may explain why our findings support threat rigidity while some prior studies do not \cite{argote1989centralize,gladstein1985group,harrington2002threat}. We also find that threat rigidity in the context of crowds appears to be much more complex than what we would expect to find in traditional groups. Nonetheless, this study presents a distinct opportunity to extend the research on threat rigidity in a more natural setting. 

Finally, the results of this study have implications for design. The literature on threat rigidity suggests that there is not one correct way for groups to respond to a shock. Therefore, systems should be designed to support sudden changes because these are likely to fluctuate with exogenous shocks. Results of our study demonstrate that crowd size, the severity of the shock, and the availability of newcomers are key factors that designers have to consider when designing systems to support crowds.

Compared to other groups, crowds operate in uniquely volatile environments where coordination is difficult and conflict is probable. We examine the impact of an external shock on crowds by analyzing the effects of the 2005 Chinese government block of Chinese Wikipedia. This event provides a natural experiment that allows us to systematically analyze the effects of a real external shock on real crowds. We find compelling evidence that both generally supports threat rigidity and contextualizes it to crowds. Our findings can help to inform both theory and  design of crowdsourcing systems. 

\section{Acknowledgements}
This research was partly supported by the National Science
Foundation under Grant No. IIS-1617820.

\bibliography{chinesewiki}
\bibliographystyle{aaai}
\end{document}